\documentclass[pr, twocolumn, superscriptaddress]{revtex4}  
\usepackage{amssymb}
\usepackage{amsmath}
\usepackage{bbm}
\usepackage{graphicx}
\usepackage{color}

\begin{document}

\author{Matthias Murray}
\affiliation{Department of Physics, University of Fribourg, 1700 Fribourg, Switzerland}
\author{Hiroshi Shinaoka}
\affiliation{Department of Physics, Saitama University, Saitama 338-8570, Japan}
\author{Philipp Werner}
\affiliation{Department of Physics, University of Fribourg, 1700 Fribourg, Switzerland}

\title{Nonequilibrium diagrammatic many-body simulations with quantics tensor trains}

\date{\today}

\begin{abstract}
The nonequilibrium Green's function formalism provides a versatile and powerful framework for  
numerical studies of nonequilibrium phenomena in correlated many-body systems. For calculations starting from an equilibrium initial state, a standard approach consists of discretizing the Kadanoff-Baym contour and implementing a causal time-stepping scheme in which the self-energy of the system plays the role of a memory kernel. This approach becomes computationally expensive at long times, because of the convolution integrals and the large amount of computer memory needed to store the Green's functions.  
A recent idea for the compression of nonequilibrium Green's functions is the quantics tensor train representation. Here, we explore this approach by implementing equilibrium and nonequilibrium simulations of the two-dimensional Hubbard model with a second-order weak-coupling approximation to the self-energy.  
We show that calculations with compressed two-time functions are possible without any loss of accuracy, and that the quantics tensor train implementation shows a much improved scaling of the computational effort and memory demand with the length of the time contour. 
\end{abstract}

\maketitle

\hyphenation{develop-ment}


\section{Introduction}

Studies of nonequilibrium phenomena in lattice systems are stimulated by experiments on laser driven solids~\cite{Giannetti2016} and cold atomic gases in modulated optical lattices~\cite{Sensarma2010}, as well as fascinating new theoretical concepts like prethermalization \cite{Berges2004} or nonthermal fixed points~\cite{Tsuji2013}. Theoretical and numerical investigations are often based on the nonequilibrium Green's function formalism~\cite{Stefanucci2013}, which provides a versatile framework and direct access to experimentally relevant probes. If the initial state of the system is an equilibrium state, the Green's functions are defined on the so-called Kadanoff Baym (KB) contour, which runs from time $0$ to some time $t_\text{max}$ along the real-time axis, returns to time $0$, and then extends to time $-i\beta$ along the imaginary-time axis (where $\beta=1/T$ is the inverse temperature of the initial state)~\cite{Aoki2014}. The interacting lattice Green's function $G_k$ for momentum $k$ is then the solution of the Dyson equation $G_k =G_{k}^0 + G_{k}^0 * \Sigma_k * G_k$, where $G_{k}^0$ is the noninteracting lattice Green's function, $\Sigma_k$ is the self-energy and ``$*$" denotes a convolution on the KB contour. In weak-coupling perturbation theories, $\Sigma_k$ is expressed diagrammatically in terms of  $G_{k}^0$ or $G_{k}$ and its calculation may require additional convolutions.   

Numerical calculations typically employ a discretization of the KB contour and a time-stepping scheme which starts from the initial equilibrium solution (imaginary-time branch) \cite{Bonitz2010,Eckstein2010}. Such nonequilibrium Green's function calculations can be conveniently implemented with high-order integration schemes using, e.~g., the NESSi library~\cite{Nessi}. A drawback of the approach is however the rapid increase with $t_\text{max}$ of the numerical cost for the calculation of the convolutions ($\sim t_\text{max}^3$), and the large amount of computer memory needed for storing two-time or higher-order Green's functions on a fine time grid ($\sim t_\text{max}^n$ for $n$ point functions). 

Various strategies have been adopted to address these challenges. One possibility is to resort to approximate schemes, like the Generalized Kadanoff-Baym Ansatz \cite{Lipavsky1986}, in which the two-time Green's function is approximately reconstructed from the density matrix. This approach has enabled nonequilibrium lattice simulations for realistic systems \cite{Schueler2020}, and there has been significant recent progress in the development of GKBA implementations with linear $t_\text{max}$ scaling~\cite{Schluenzen2020,Pavlyukh2022}. A more controlled approximation, which works well if the self-energy decays fast away from the diagonal $t=t'$, is the truncation of the memory time in $\Sigma_k(t,t')$ \cite{Schueler2018}. In this case the convolutions don't need to be performed over the full KB contour, but only over some time interval defined by the cutoff time $t_\text{cut}$, and also the storage requirement is reduced \cite{Stahl2022}. 

A recent and promising idea, which avoids any approximations, is to apply memory compression techniques to the nonequilibrium Green's functions. Ref.~\onlinecite{Kaye2021} combined 
a hierarchical low-rank representation of the Green's function with a time-stepping scheme and demonstrated a memory reduction from $\mathcal{O}(t_\text{max}^2)$ to $\mathcal{O}(t_\text{max})$ and an improved scaling in the solution of Dyson equations. This innovation allows to time-propagate nonequilibrium Green's function calculations to $t_\text{max}$ which would be inaccessible without compression. In a separate development, quantics tensor train (QTT) representations of multi-variable functions were introduced in Ref.~\onlinecite{Shinaoka2023} and shown to enable high compression ratios for typical nonequilibrium Green's functions. This approach in principle enables a simultaneous compression of the time and space (or momentum) dependence of nonequilibrium Green's functions. In the context of diagrammatic many-body calculations, it is however useful only if the entire simulation, including the evaluation of the self-energy and the solution of Dyson equations, can be implemented in compressed form. 

In this paper, we provide a proof-of-principles for diagrammatic calculations based on QTT compressed nonequilibrium Green's functions by implementing self-consistent second-order  perturbative solutions of the two-dimensional (2D) Hubbard model, both for equilibrium and nonequilibrium setups. We employ Green's functions on the unfolded KB contour and focus on the compression of the time-dependence. We explain the implementation of the various steps in the diagrammatic calculation and discuss the memory requirement and efficiency of our implementation.  

The paper is organized as follows. In Sec.~\ref{sec:method} we describe the model studied and the methodology. Section~\ref{sec:results} presents test results for the solution of the equilibrium and quenched 2D Hubbard model, while Sec.~\ref{sec:conclusions} is a short conclusion.  

\begin{figure}[t]
  \includegraphics[width=0.9\linewidth]{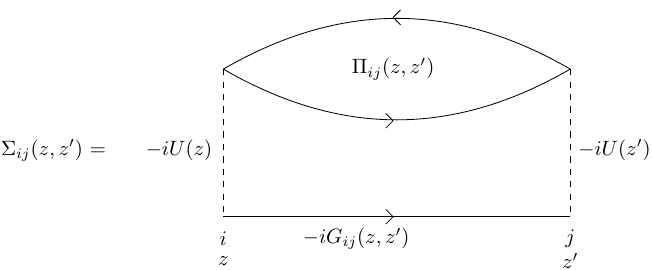}
  \caption{Second-order contribution to the self-energy in the real-space representation. $z$ and $z'$ are time points on the KB contour $\mathcal{C}$.}
\label{fig_sigma}
\end{figure}

\section{Formalism}
\label{sec:method}

\subsection{Model and second-order perturbation theory}

We consider the half-filled 2D Hubbard model on a square lattice. The Hamiltonian is 
\begin{align}
H(t) =& -v \sum_{\langle ij\rangle\sigma} c^\dagger_{i\sigma}c_{j\sigma} + U(t)\sum_i (n_{i\uparrow}-\tfrac 12)(n_{i\downarrow}-\tfrac 12)
\end{align}
with $c^\dagger_{i\sigma}$ the creation operator for an electron with spin $\sigma$ on site $i$, $v$ the nearest-neighbor hopping, and $U$ the on-site interaction (which in the quench calculation depends on time $t$). In the first term, $\langle ij\rangle$ denotes nearest-neighbor sites. The dispersion of the noninteracting 2D model is $\epsilon_k=-2v(\cos k_x + \cos k_y)$, where we set the lattice constant $a$ to unity. In the rest of the paper, we use $v=1$ as the unit of energy ($\hbar/v\equiv 1/v$ as the unit of time). We furthermore suppress the spin index, since we will restrict the calculations to paramagnetic states.

As a simple but nontrivial example of a diagrammatic calculation, we consider self-consistent second order perturbation theory, corresponding to the real-space self-energy illustrated in Fig.~\ref{fig_sigma}. Introducing the polarization bubble 
\begin{equation}
\Pi_{ij}(z,z')=-G_{ij}(z,z')G_{ji}(z',z)
\label{eq_pi}
\end{equation}
formed by the interacting lattice Green's functions $G_{ij}(z,z')=-i\langle \mathcal{T}_\mathcal{C}c_i(z)c^\dagger_j(z')\rangle$ ($\mathcal{T}_\mathcal{C}$ is the time ordering operator on the KB contour $\mathcal{C}=\mathcal{C}_1\bigcup \mathcal{C}_2\bigcup \mathcal{C}_3$ and $z$ denotes the contour time), we can express the self-energy as
\begin{equation}
\Sigma_{ij}(z,z')=U(z)G_{ij}(z,z')\Pi_{ij}(z,z')U(z').
\label{eq_sigma}
\end{equation}
Fourier transformation of the space-translation invariant functions $G_{ij}$ and $\Sigma_{ij}$ to momentum space ($f(k)=\sum_r e^{-ik r}f(r)$, $f(r)=\frac{1}{N_k^2}\sum_k e^{ikr}f(k)$, $N_k^2$ denotes the total number of sites or momentum points) yields $G_k(z,z')$ and $\Sigma_k(z,z')$. The noninteracting Green's function is determined by the dispersion $\epsilon_k$ and the Fermi function $f_T$ for the initial temperature \cite{Aoki2014},
\begin{equation}
G^0_k(z,z')=-i[\theta_\mathcal{C}(z,z')-f_T(\epsilon_k(0))]e^{-i\int_{z'}^{z} d\bar z \epsilon_k(\bar z)},
\end{equation}
where $\theta_\mathcal{C}(z,z')$ is the step function defined on the KB contour. With these, we can solve the lattice Dyson equation
\begin{equation}
G_k(z,z') =G_{k}^0(z,z') + [G_{k}^0 * \Sigma_k * G_k](z,z')
\label{eq_dyson}
\end{equation}
to obtain an updated lattice Green's function $G_k$, which can then be Fourier transformed to real space and used to compute an updated self-energy. The whole procedure is  iterated until convergence is reached. 

\begin{figure}[t]
  \includegraphics[width=0.9\linewidth]{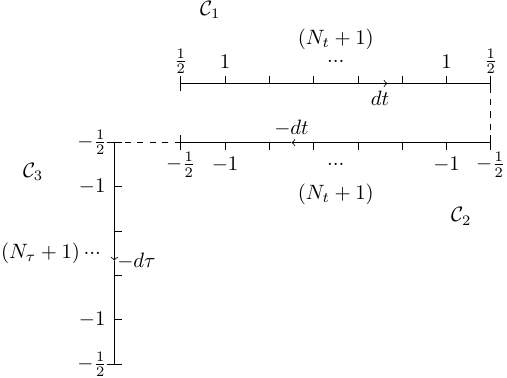}
  \caption{Discretization of the three-legged KB contour $\mathcal{C}$ and weight factors for the trapezoidal rule integration.
  }
\label{fig_KB}
\end{figure}

\subsection{Discretized KB contour and matrix formulation}
\label{sec_discretized_KB}

\begin{figure}[t]
\centering
\begin{minipage}{0.5\textwidth}
  \includegraphics[width=1.05\linewidth]{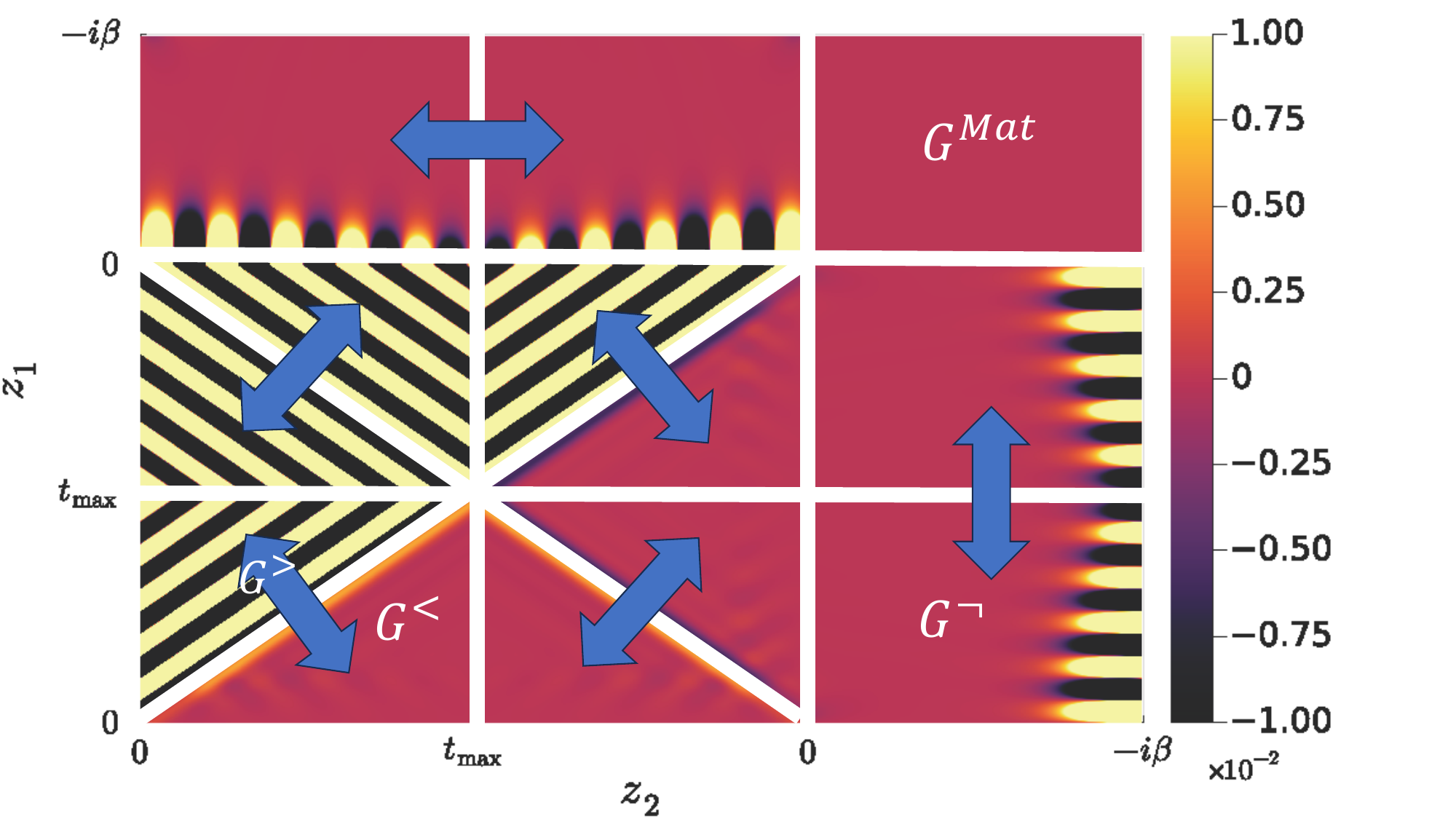}
\end{minipage}\\
\begin{minipage}{0.5\textwidth}
   \includegraphics[width=1\linewidth]{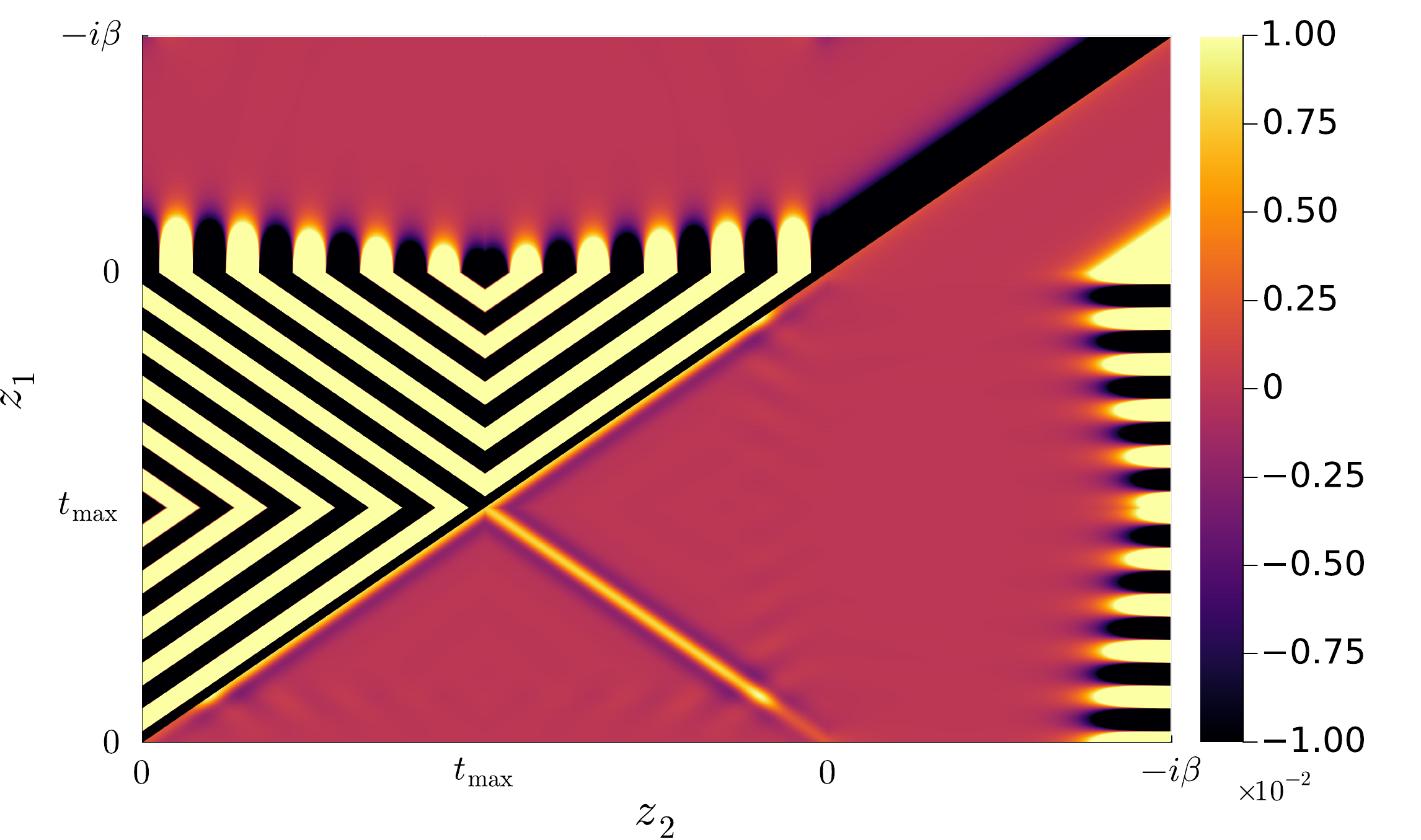}
\end{minipage}
  \caption{Equilibrium unfolded Green's function $G_k(z,z')$ for $U=2$, $\beta=5$, $N_t=N_{\tau}=800$ and $k=(0, 0)$. The top (bottom) panel shows the real (imaginary) part.
  In the top panel we also indicate the different Green's function components and symmetry relations with respect to different axes.
  }
\label{fig_matrix}
\end{figure}

We first discuss a simple and straight-forward strategy for solving Eqs.~\eqref{eq_pi}, \eqref{eq_sigma} and \eqref{eq_dyson}, which relies on the discretization of the KB contour and the matrix representation of $G$, $\Sigma$ and $\Pi$. We illustrate the discretized
contour in Fig. 2. The forward and backward branches $\mathcal{C}_1$ and $\mathcal{C}_2$ are represented by $(N_t+1)$ grid points with a spacing of $dt=t_\text{max}/N_t$, while the Matsubara branch $\mathcal{C}_3$ is represented by $(N_\tau+1)$ grid points with a spacing $d\tau=\beta/N_\tau$ ($\beta$ is the inverse temperature). 
In Fig.~\ref{fig_matrix} we plot a typical example of an unfolded $G$ in the space of $z$ and $z'$. In the real part (top panel), we also indicate the greater ($G^>$), lesser ($G^<$), left-mixing ($G^\neg$) and Matsubara ($G^\text{Mat}$) components, which determine the whole matrix via symmetry operations that can be easily deduced from the color map, and which are indicated by the blue arrows. (To better reveal the structures, the color bar is limited to the range $[-0.01,0.01]$.) The function shown corresponds to the equilibrium solution for $k=(0,0)$, $U=2$, $\beta=5$ and to a time-grid with $N_t=800$ discretization steps on the real-time axis and $N_\tau=800$ steps on the Matsubara axis. There are thus a total of $2403$ points on the unfolded KB contour. Storing such a Green's function with $2403^2$ complex numbers requires $88.1$~MB of memory. With the Fourier transformed unfolded Green's functions, $\Pi_{ij}$ and $\Sigma_{ij}$ can be calculated by element-wise products.

In the Dyson equation \eqref{eq_dyson} one also needs to take into account the direction of the time-integral in the convolutions. In the discretized convolution integrals, this can be done by introducing the diagonal matrix $\tau(z,z')=\text{diag}(dt/2$, $dt$, $\dots$, $dt$, $dt/2$, $-dt/2$, $-dt$, $\ldots$, $-dt$, $-dt/2$, $-i d\tau/2$, $-i d\tau$, $\ldots$, $-i d\tau$, $-i d\tau/2)$, corresponding to the trapezoidal integration rule. The weight factors associated with the different grid points are illustrated in Fig.~\ref{fig_KB}. With this, the Dyson equation becomes the matrix equation 
\begin{equation}
\underline G_k =\underline G_{k}^0 + \underline G_{k}^0 * \underline  \tau * \underline \Sigma_k * \underline \tau * \underline G_k, \label{eq_dyson_matrix}
\end{equation}
where we denote the matrices in the discretized $(z,z')$ space by an underline and the star symbols here represent matrix multiplications. 
In practice, it may be convenient to combine the (possibly time-dependent) interaction $U(z)$ and $\tau(z,z')$ into the diagonal matrix $U_\tau(z,z')=\text{diag}(U(0)dt/2,U(dt)dt,\dots )$  
and to pull the $U$-factors out of Eq.~\eqref{eq_sigma}.

The solutions obtained with these discretized functions and matrix equations will serve as a reference for the quantics tensor train implementation discussed in the next section.

\subsection{Implementation with quantics tensor trains}
\label{sec_implementation}

\subsubsection{Tensor train representation of two-time functions}

\begin{figure}[t]
\begin{center}
\includegraphics[angle=0, width=\columnwidth]{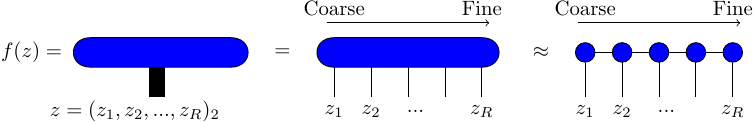}
\caption{
Illustration of the quantics tensor train representation of the function $f(z)$. First, the argument $z$ is expressed in a binary representation, which introduces the bits (or spins) $z_i$, $i=1, \ldots, R$. Then, the function defined on the $2^R$ dimensional space is decomposed into a tensor train. 
}
\label{fig_tensortrain}
\end{center}
\end{figure}

A general strategy for compressing (multi-variable) functions is the QTT representation, which was recently presented and analyzed in the context of many-body calculations in Ref.~\onlinecite{Shinaoka2023}.
We first briefly discuss the main idea for a function $f(z)$ which depends on a single variable $z$ defined on the interval $[0,z_\text{max}]$. Let us divide the time-interval into $N_z=2^R-1$ slices of length $d z$ ($2^R$ time points) and map the discretized times to binary numbers $(z_1,\ldots,z_R)_2$ representing these grid points:  $(0, \ldots, 0)_2$ corresponds to the first grid point $z=0$ and $(1,\ldots, 1)_2$ to the last grid point $z=z_\text{max}=N_z d z$. 
Physically, this procedure can be thought of as mapping the discretized time interval onto the $2^R$ dimensional Hilbert space of a spin-1/2 system. The function $f$ defined on this space may now be represented as a tensor train (or matrix product state \cite{Schollwock2011,Cirac2021}), as illustrated in Fig.~\ref{fig_tensortrain}. Here, the bond dimension $D$ of the tensors is controlled by a parameter $\epsilon_\text{cutoff}$, which defines a cutoff in the singular values retained in the construction of the tensor train. Specifically, we measure the accuracy with respect to the Frobenius norm $|\cdots |$ as
\begin{align}
\epsilon_\mathrm{cutoff} &= \frac{|A - \tilde A |^2}{|A|^2},
\end{align}
where $A$ is the original tensor or MPS, and $\tilde A$ is the truncated MPS.
We refer the reader to Appendix A of Ref.~\onlinecite{Shinaoka2023} for a more detailed description.

The approach can be extended to multi-variable functions, such as the two-time Green's function $G(z,z')$ or self-energy $\Sigma(z,z')$, by arranging the corresponding digits of the binary representations of $z=(z_1,\ldots,z_R)_2$ and $z'=(z'_1,\ldots,z'_R)_2$ into the bit string $(z_1 , z'_1, \ldots ,z_R ,z'_R)_2$ with $R'=2R$ bits. In principle, the binary representation of the time variables could also be combined with binary representations of the space or momentum variables, but in the present study, we will restrict ourselves to the quantics representation of the (contour) time variables. 

It has been shown in Ref.~\onlinecite{Shinaoka2023} that generically, for reasonable values of $\epsilon_\text{cutoff}$, the scale separation inherent to most physical functions leads to three distinct regimes in the evolution of the bond dimension along the tensor train. First, the bond dimension increases exponentially, then reaches a plateau in the region associated with intermediate scales, and eventually decreases since the behavior on very short scales is often associated with noise and lacks relevant information. As a result of this structure, the tensor train representation enables a significantly compressed representation of the function, compared to the original one on the discrete time grid, with a practically negligible loss of accuracy.  

\subsubsection{Diagrammatic calculations with tensor trains}
\label{sec_implementation}

In order to perform diagrammatic calculations like the second-order solution of the Hubbard model with compressed objects, we must implement the relevant steps in these calculations with quantics tensor trains. These steps are (i) Fourier transformations, (ii) the calculation of element-wise products, as in Eq.~\eqref{eq_pi} with constant $U$, (iii) the multiplication with scalars, as in Eq.~\eqref{eq_sigma}, and the calculations of (iv) sums and (v) convolutions, as in Eq.~\eqref{eq_dyson}. In the following, we briefly explain the implementation of these fundamental operations.

\paragraph{Multiplication with scalar.}
Let $\widehat{f}(z_1,\ldots,z_R)=\widehat{f}^{(1)}(z_1)\cdot\ldots\cdot\widehat{f}^{(R)}(z_R)$ be a QTT representation of $f(z)$. 
Here, $\widehat{f}^{(j)}(z_j)$ represents an individual tensor and the dot symbols indicate tensor products. 
To perform a multiplication with a scalar $a$ in the QTT representation, we can multiply any single one of the $R$ tensors: $a \widehat{f}(z_1,\ldots,z_R)=[a \widehat{f}^{(1)}(z_1)]\cdot\ldots\cdot\widehat{f}^{(R)}(z_R)=\ldots=\widehat{f}^{(1)}(z_1)\cdot\ldots\cdot[a\widehat{f}^{(R)}(z_R)]$.
This operation does not change the bond dimensions of the QTT. 

\paragraph{Sum.}
A naive approach to sum two QTTs $\widehat{f}_1$ and $\widehat{f}_2$, with maximum bond dimensions $D_1$ and $D_2$, 
respectively, is to make use of direct sums of the two underlying spaces. For $\widehat{f} = \widehat{f}_1 + \widehat{f}_2 \equiv \widehat{f}^{(1)}(z_1)\cdot\ldots\cdot\widehat{f}^{(R)}(z_R)$, this would result in 
\begin{align}
    \widehat{f}^{(j)}(z_j) =  \widehat{f}_1^{(j)}(z_j)\bigoplus \widehat{f}_2^{(j)}(z_j).
\end{align}
For example, for $j=1$ $(j=R)$, the tensors are simply matrices, which means that we concatenate the two columns (rows) of each site. This can however lead to much redundancy, as the resulting maximum bond dimension is $D=D_1+D_2$. To see this, consider the case $\widehat{f}_2=\widehat{f}_1$, where this approach leads to $D=2D_1$. On the other hand, this sum is the same as a multiplication by a factor $2$, where the latter operation keeps the maximum bond dimension at $D=D_1$. After a sum, it is thus necessary to re-compress the resulting QTT to a lower-rank representation \cite{Schollwock2011}. The number of operations for the sum scales as $\mathcal{O}((D_1+D_2)^3)$ \cite{Shinaoka2023}.

\paragraph{Fourier transformation.} 
Let $f_{r}(z)$ be functions of $z$, where $r$ is defined on a mesh of size $N^2=N_k^2$. The Fourier transform with respect to $r$ of its QTT representation $\widehat{f}_{r} (z_1,\ldots,z_R)$ is given by 
\begin{equation}
    \widehat{f}_{k}(z_1,\ldots,z_R) = \sum_{r} e^{ikr}\ \widehat{f}_{r}(z_1,\ldots,z_R),
\end{equation}
which can be simply implemented as the sum over QTTs multiplied by scalars. 
Here, we use a naive approach for the Fourier transform. For large $N_k$, it may be beneficial to combine the Fast Fourier Transform (FFT) algorithm with QTTs.

\paragraph{Element-wise product.} 
To perform an element-wise multiplication of two QTTs $\widehat{f}_i(z)=\widehat{f}_i^{(1)}(z_1)\cdot\ldots\cdot\widehat{f}_i^{(R)}(z_R)$, $i=1,2$, we transform the first one into a higher rank diagonal representation \cite{Shinaoka2023}
\begin{align}
    \widehat{\overline{f}}_1(z_1,z'_{1},\ldots,z_R,z'_{R})=&(\widehat{f}_1^{(1)}(z_1)\ \delta_{z_1,z'_1})\cdot\ldots\notag\\
    &\ldots\cdot(\widehat{f}_1^{(R)}(z_R)\ \delta_{z_R,z'_R}).
\end{align}
Then, the contraction over common indices
\begin{equation}
    \sum_{z'_{1},\ldots,z'_{R}}\widehat{\overline{f}}_1(z_1,z'_{1},\ldots,z_R,z'_{R})
    \ \widehat{f}_2(z'_1,\ldots,z'_R)
\end{equation}
yields the desired result. A naive implementation would lead to an inefficient scaling $\mathcal{O}(D^6)$. Fortunately, in practice, it is possible to reduce this to $\mathcal{O}(D^4)$ (see Fig.~25(b) in Ref.~\cite{Shinaoka2023})
by making use of a fitting algorithm with a two-site update for the contraction~\cite{Stoudenmire2010}.

\paragraph{Convolution.}
Let $f_1(z,z')$ and $f_2(z,z')$ be two-time functions defined on the KB contour. As discussed in Sec.~\ref{sec_discretized_KB}, the contour convolution $\int_{\mathcal{C}} d\overline{z}\ f_1(z, \overline{z})\ f_2(\overline{z},z')$ can be implemented as the matrix multiplication $\underline f_1 * \underline \tau * \underline f_2$, with $\underline \tau$ a diagonal matrix. It thus corresponds to two matrix multiplications. Here, we explain how to implement a single matrix multiplication corresponding to $\underline f_1 * \underline f_2$.  The contraction \cite{Shinaoka2023}
\begin{align}
    \sum_{\overline{z}_1, \overline{z}'_1,\ldots, \overline{z}_R, \overline{z}'_R} &\widehat{\overline{f}}_1(z_1, z'_1, \overline{z}_1, \overline{z}_1',\ldots,z_R, z'_R, \overline{z}_R, \overline{z}_R')\notag \\
    &\times\widehat{f}_2(\overline{z}_1, \overline{z}_1',\ldots,\overline{z}_R, \overline{z}_R')
\end{align}
of QTTs represents this matrix multiplication in compressed form. Here, $\widehat{f}_2$ is the QTT corresponding to $f_2$ and $\widehat{\overline{f}}_1$ is an auxiliary QTT with new combined indices on each site. Concretely, this can be done by first contracting each pair of neighboring sites (of both QTTs) and then contracting over the ``column" and ``row" indices of the resulting QTTs. We refer to section III~\!C in Ref.~\onlinecite{Shinaoka2023} for a detailed description. This operation again scales as $\mathcal{O}(D^4)$ \cite{Shinaoka2023} if the fitting algorithm \cite{Stoudenmire2010} is used.

\section{Results}
\label{sec:results}

\subsection{Compressibility of $G_k$ and $\Sigma_k$}

\begin{figure}[t]
\centering 
\advance\leftskip-4.6cm
\begin{minipage}{0.5\linewidth}
      \includegraphics[width=2.05\linewidth]{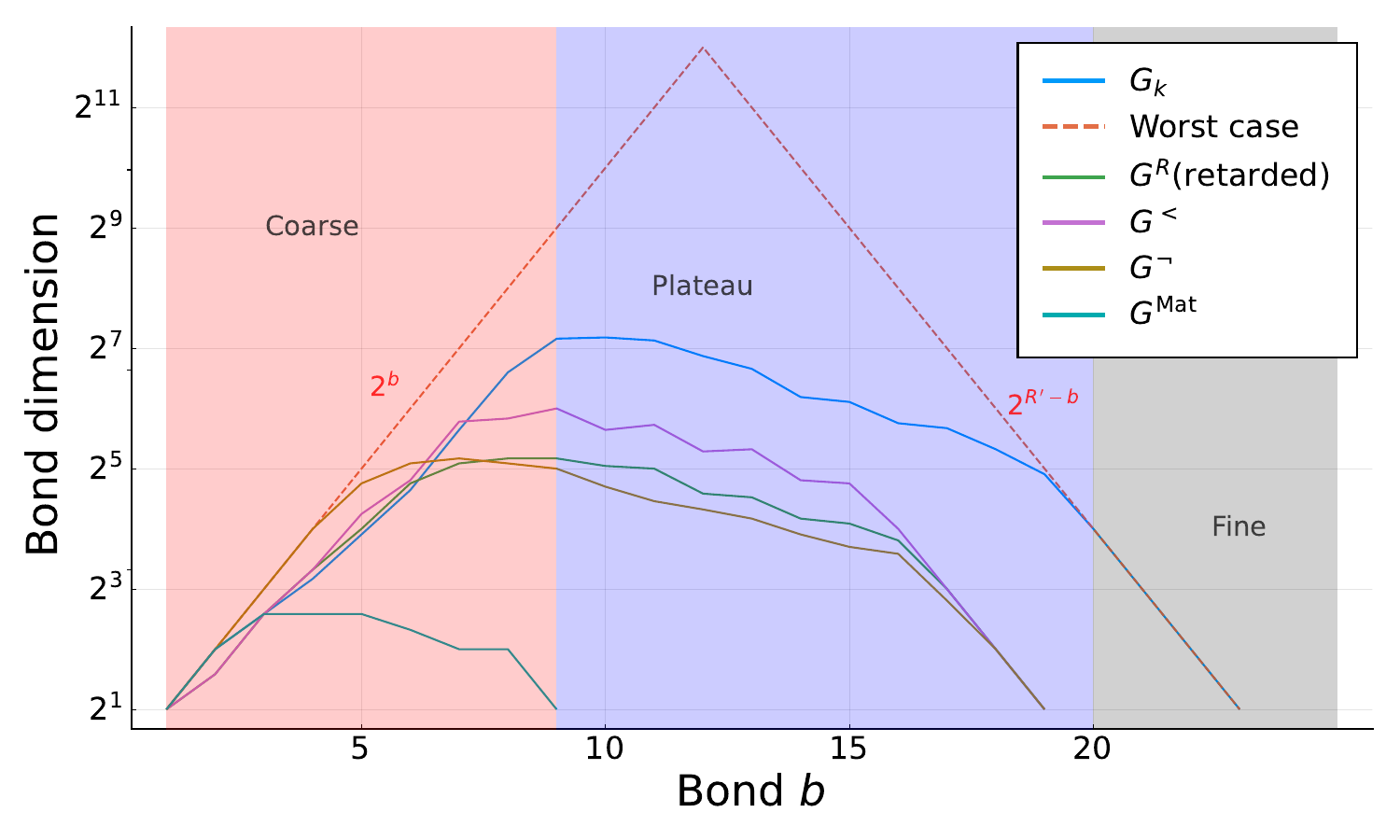}
\end{minipage}\\
\begin{minipage}{0.5\linewidth}
   \includegraphics[width=2.05\linewidth]{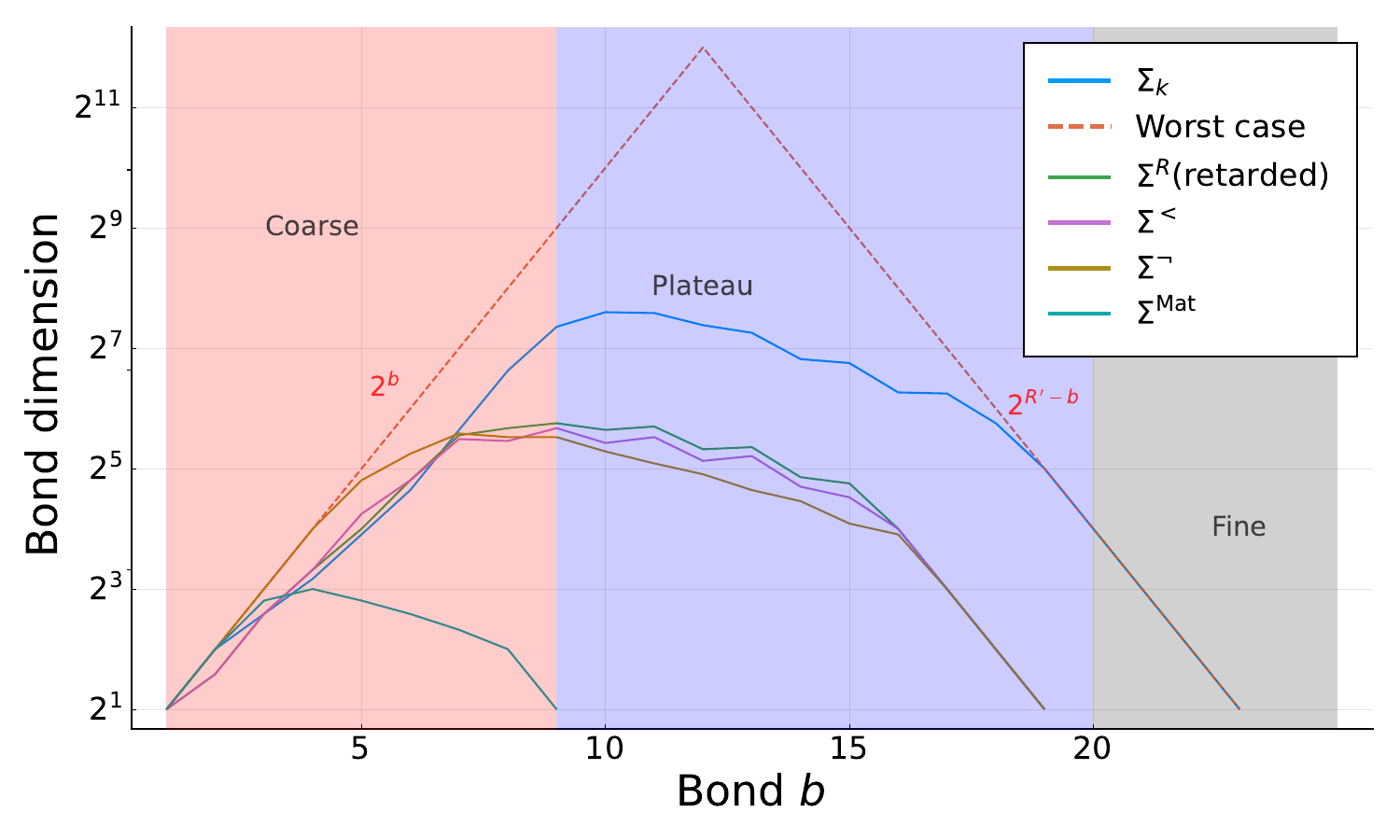}
\end{minipage}
  \caption{Bond dimensions of the Green's function (top) and self-energy (bottom) on a logarithmic scale, as well as the results for the individual components, for cutoff $\epsilon_\text{cutoff} = 10^{-15}$. The parameters are  $k = (0, 0)$, $U=2$, $\beta=5$, $N_t =N_{\tau} = 800$. The QTT has $R' = 2R = 24$ bits ($R=12$ for each time variable) in case of the full function, $20$ bits for the retarded, lesser, and left-mixing components, and $10$ bits in case of the Matsubara component. 
  The dashed line represents the worst case scenario 
  without scale separation.}
  \label{fig_bonddimension}
\end{figure}

To investigate the compressibility of typical momentum-dependent Green's functions and self-energies, we consider the equilibrium solutions for $U=2$, inverse temperature $\beta=5$ and $N_t, N_\tau=800$. In Fig.~\ref{fig_bonddimension} we plot the bond dimensions of the tensor train representation of the $k=(k_x,k_y)=(0,0)$ Green's function and self-energy, both for the functions defined on the unfolded KB contour (similar to Fig.~\ref{fig_matrix}) and for the individual components (lesser, retarded, left-mixing and Matsubara). 
Here we use $\epsilon_\text{cutoff}=10^{-15}$, which assures a highly accurate QTT representation of the original functions.

Focusing first on the results for the unfolded contour, where the functions contain cusps and discontinuities, as well as redundant parts, we observe an exponential increase in the bond dimension up to a value of about 150 at the 9th link. This is followed by a rough ``plateau", and eventually an exponential decrease in the bond dimensions. These bond dimensions correspond to a compression ratio (ratio of the memory needed to store the QTT and matrix representation) of $0.0135$ for the Green's function and $0.0240$ for the self-energy. 

As shown in the same plots, the bond dimensions for the tensor train representations of the individual components are considerably smaller, and the plateau appears earlier. Nevertheless, because there is no redundant information if we consider the components, the compression ratios are not very different than for the full functions: In the case of the Green's function, the results in Fig.~\ref{fig_bonddimension} correspond to the compression ratios $0.0409$ (lesser), $0.0174$ (retarded) and $0.0160$ (left-mixing). The corresponding values for the self-energy are $0.0302$ (lesser), $0.0364$ (retarded) and $0.0275$ (left-mixing).

For the efficiency of the diagrammatic calculation in the QTT form, the maximum bond dimension $D$ is crucial (see Sec.~\ref{sec_implementation}). Hence, even though the QTT representation can reproduce functions with cusps and discontinuities up to machine precision \cite{Shinaoka2023}, these result show that an efficient implementation of diagrammatic calculations should make use of compressed components and Langreth rules \cite{Langreth1976}, rather than the functions defined on the unfolded KB contour. More specifically, with 4 independent components and a maximum bond dimension of $2^x$ for these components, the maximum bond dimension of the full unfolded Green's function or self-energy can be estimated to be approximately $4\cdot 2^x=2^{2+x}$. This roughly explains the higher maximum bond dimension of the functions defined on the unfolded contour in Fig.~\ref{fig_bonddimension} ($x\approx 5.5$ in the case of $\Sigma$, maximum bond dimension $\approx 2^{5.5}$ for the components and $\approx 2^{7.5}$ for the full functions). Nevertheless, for the current proof-of-principle calculations, we will proceed with compressed two-time functions defined on the unfolded KB contour.   

\begin{figure}[t]
\begin{minipage}{0.49\linewidth}
  \includegraphics[width=1\linewidth]{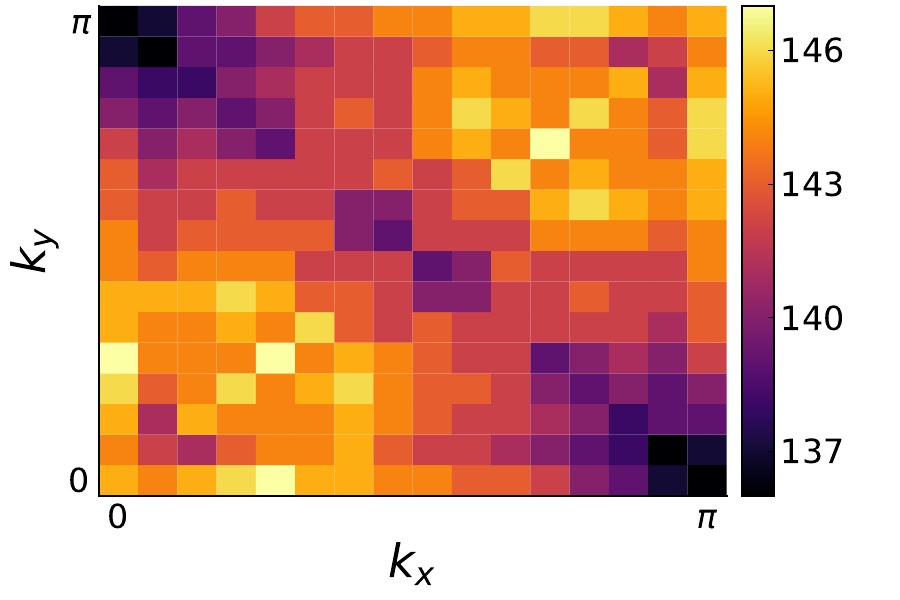}
\end{minipage}
\begin{minipage}{0.49\linewidth}
    \includegraphics[width=1\linewidth]{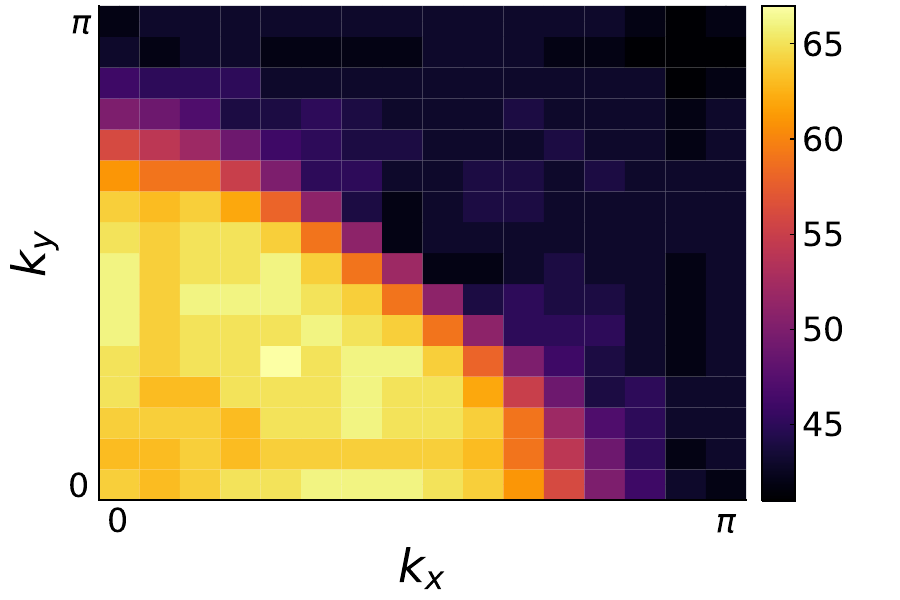}
\end{minipage}
\begin{minipage}{0.49\linewidth}
       \includegraphics[width=1\linewidth]{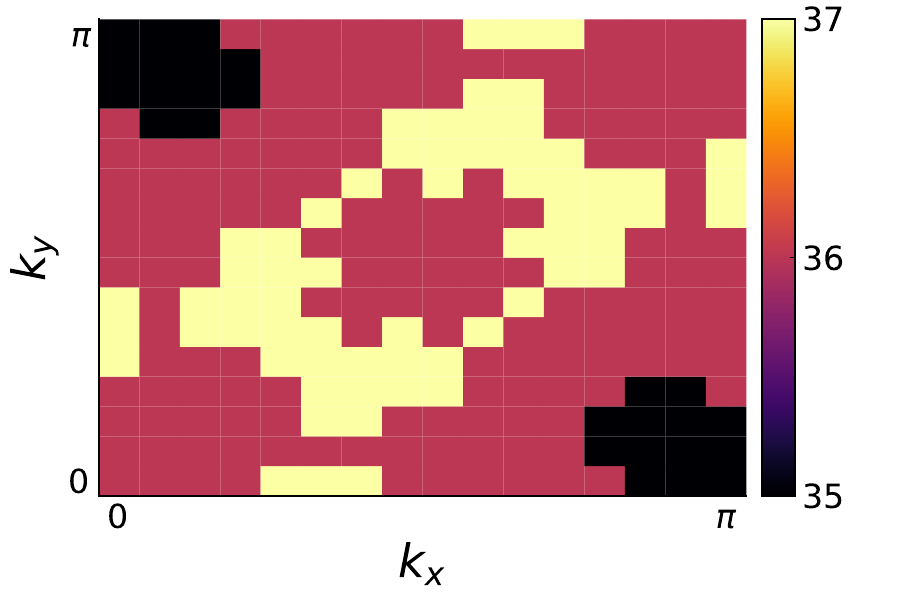}
\end{minipage}
\begin{minipage}{0.49\linewidth}
  \includegraphics[width=1\linewidth]{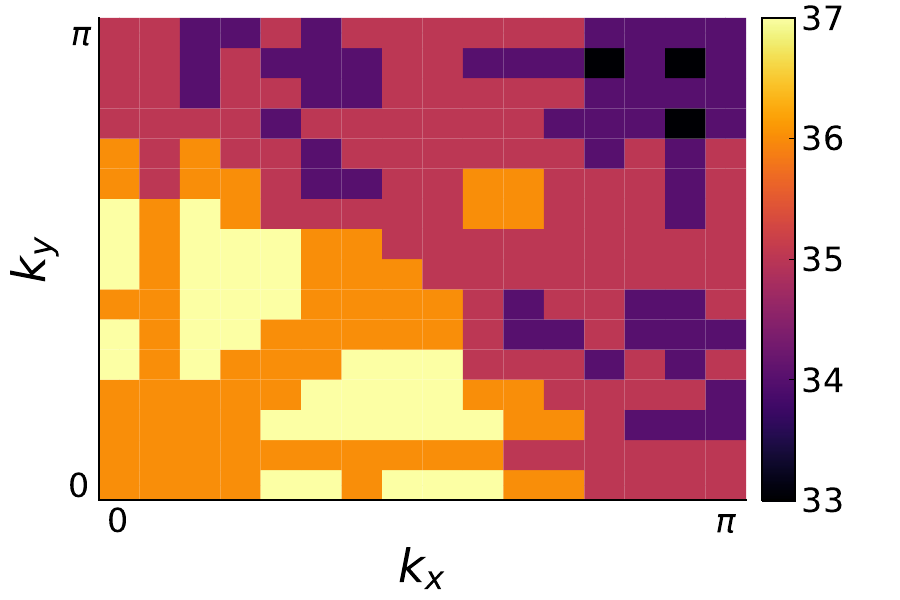}
\end{minipage}
  \caption{Maximum bond dimension as a function of $k$ in a quarter of the first BZ for the full Green's function $G_k$ (top left) and its lesser (top right), retarded (bottom left), and left-mixing (bottom right) components.
  $U=2$, $\beta=5$, $N_t =N_{\tau} = 800$, $\epsilon_\text{cutoff} = 10^{-15}$.}
\label{fig_bonddimension_k}
\end{figure}

One may also wonder how the compressibility of $G_k$ depends on the momentum $k$. To illustrate this, we plot in Fig.~\ref{fig_bonddimension_k} the maximum bond dimension of the QTT within a quarter of the first Brillouin zone (BZ). The top left panel shows the results for the function defined on the unfolded contour, and the other panels for the lesser, retarded, and left-mixing components. While the variation with $k$ is not very large in the case of the full $G_k$, we find that the maximum bond dimension is lowest along the Fermi surface. In the case of the lesser component, the bond dimension is larger in the filled part of the BZ (where the lesser spectrum has a peak) than in the empty part (where the lesser spectrum is very small). In contrast, the retarded component, whose spectrum exhibits a quasi-particle peak for all $k$, has an almost constant maximum bond dimension in the entire BZ. In the case of the left-mixing component, one finds a gradual increase in the maximum bond dimension as one moves from the unoccupied to the occupied part, with a maximum bond dimension roughly half-way between the Fermi surface and the $\Gamma$ point.   

\begin{figure}[t]
\begin{minipage}{0.49\linewidth}
  \includegraphics[width=1\linewidth]{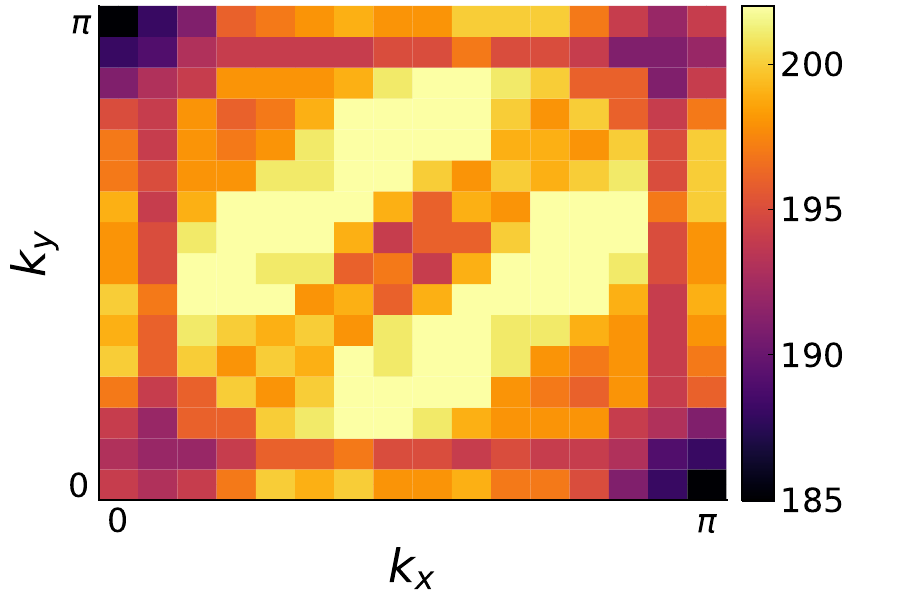}
\end{minipage}
\begin{minipage}{0.49\linewidth}
    \includegraphics[width=1\linewidth]{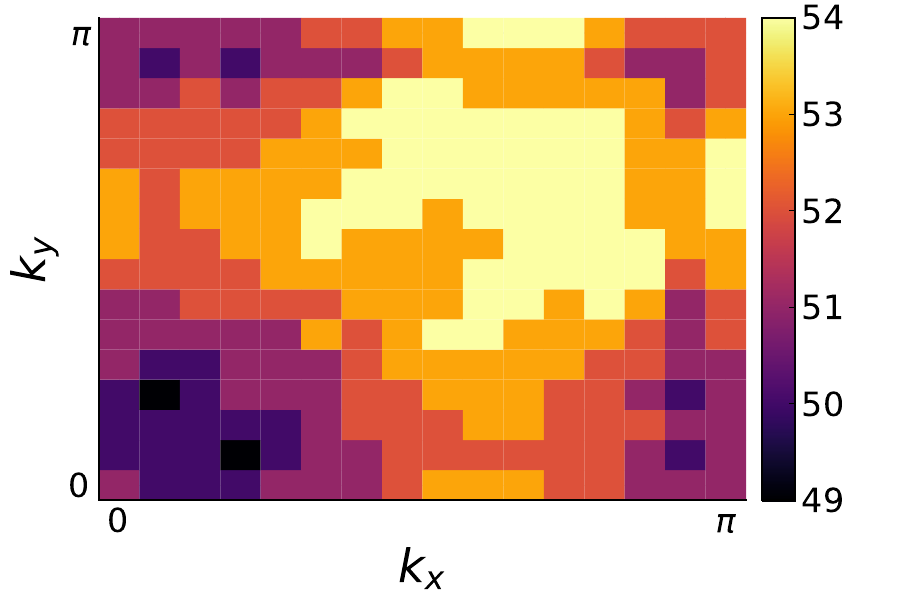}
\end{minipage}
\begin{minipage}{0.49\linewidth}
       \includegraphics[width=1\linewidth]{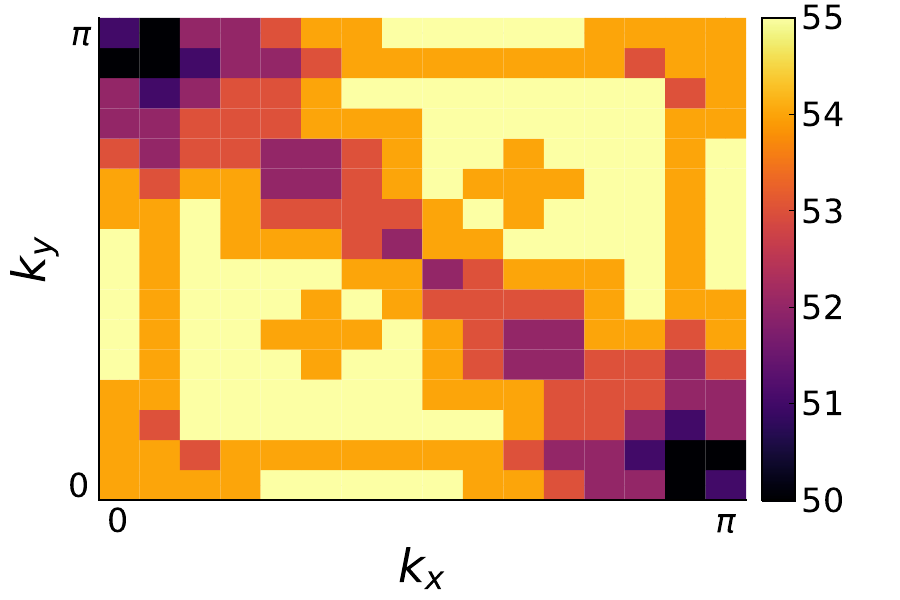}
\end{minipage}
\begin{minipage}{0.49\linewidth}
  \includegraphics[width=1\linewidth]{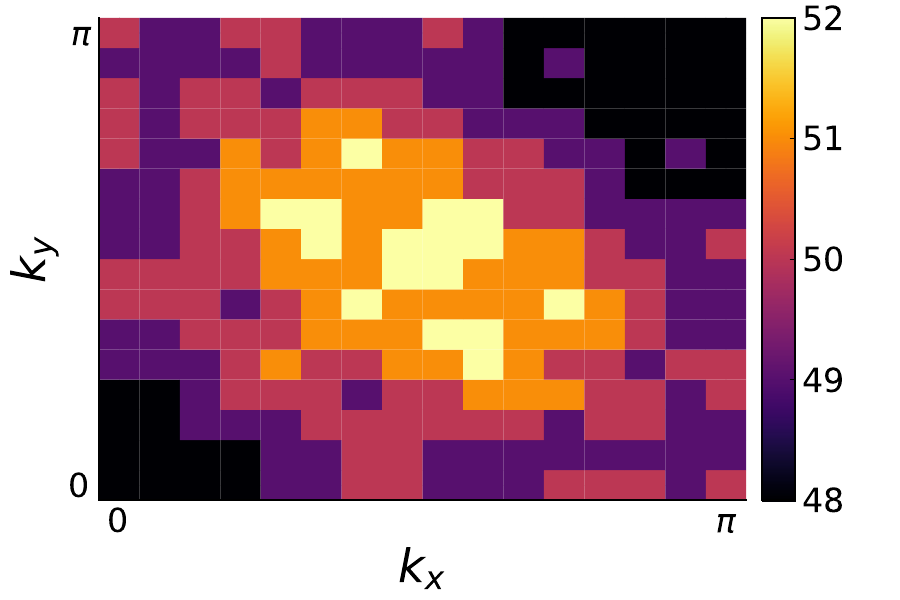}
\end{minipage}
  \caption{Maximum bond dimension as a function of $k$ in a quarter of the first BZ for the full self-energy $\Sigma_k$ (top left) and its lesser (top right), retarded (bottom left), and left-mixing (bottom right) components.
  $U=2$, $\beta=5$, $N_t =N_{\tau} = 800$, $\epsilon_\text{cutoff} = 10^{-15}$.
  }
\label{fig_bonddimension_sigma_k}
\end{figure}

The maximum bond dimensions for the self-energy and its components are plotted as a function of $k$ in Fig.~\ref{fig_bonddimension_sigma_k}. While the bond dimensions for $\Sigma_k$ are generally larger than for $G_k$, as already seen in Fig.~\ref{fig_bonddimension}, the maximum bond dimension is almost independent of $k$, even for the components. This is because the self-energy expression involves products of different Green's function components. For example, in real space, the lesser component of $\Pi$ is a product of the lesser and greater components of $G$. 

\subsection{Exponential convergence with $R$}

\begin{figure}[t]
  \includegraphics[width=1\linewidth]{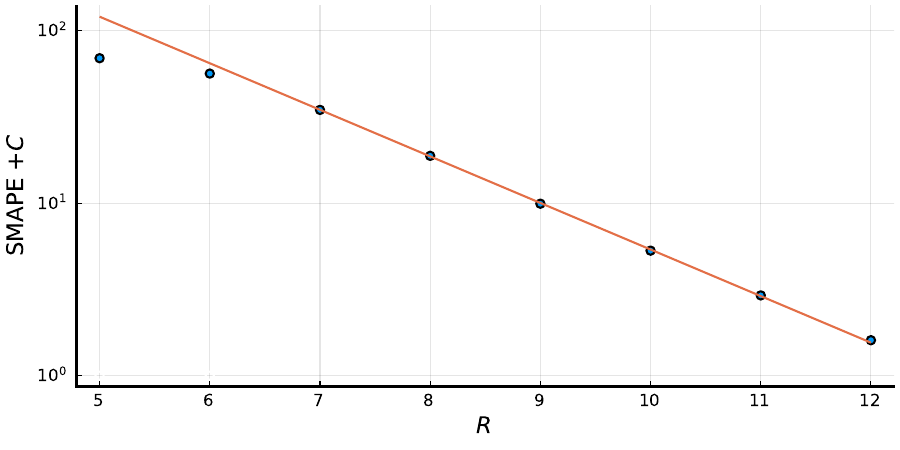}
  \caption{Exponential convergence of the QTT representation with $R$. The figure plots SMAPE for the reference state $G_\text{matrix}$ (circles). An offset $C=1.6$ is added, because the exponential convergence is towards the infinite-resolution function $G_\infty$. 
  }
  \label{fig_smape_expo}
\end{figure}
An attractive feature of the QTT compression is that the accuracy of the compressed representation, and hence time evolution, increases exponentially with increasing $R$. 
This is demonstrated in Fig.~\ref{fig_smape_expo}, where we plot the deviation between the QTT compressed Green's function with $R'=2R$ bits and the matrix representation of the Green's function for the smallest time step (largest $R$). 
The reference Green's function here is the same as in Fig.~\ref{fig_bonddimension}, i.e. the converged interacting $G_k(z,z')$ for $k = (0, 0)$, $U=2$, and $\beta=5$. 
The deviations between the Green's functions from the two methods is provided by the symmetric mean absolute percentage error (SMAPE) defined as 
\begin{equation}
\text{SMAPE}=\frac{100}{2^{2R}}\sum
_{z,z'} \frac{|G_\text{QTT}(z,z')-G_\text{matrix}(z,z')|}{|G_\text{QTT}(z,z')|+|G_\text{matrix}(z,z')|}, 
\label{eq_smape}
\end{equation}
where $G_\text{QTT}$ ($G_\text{matrix}$) is the Green's function from the QTT (matrix) implementation, and the sums are over the discretized contour.

Since the reference $G_\text{matrix}$ itself has a finite resolution (corresponding to $R=12$), we plot the SMAPE result in Fig.~\ref{fig_smape_expo} with an offset $C=1.6$, which represents the deviation to the infinite resolution Green's function $G_\infty$. 
The offset was determined by fitting the SMAPE data in the interval $7\le R\le 12$ to the function $\exp(-\alpha R)-C$, which yields $\alpha=0.62\pm 0.01$ and $C=1.6 \pm 0.1$.
The log-scale plot in Fig.~\ref{fig_smape_expo} hence shows the exponential convergence towards $G_\infty$.

\subsection{Solution of the Dyson equation}

We now use the QTT representations of $G^0_k$ and $G_k$ to construct the self-energy $\Sigma_k$ and to iteratively solve the Dyson equation \eqref{eq_dyson} using the routines described in Sec.~\ref{sec_implementation}. After the generation and compression of the $G^0_k$, we work exclusively with quantics tensor trains, and convert the results to functions on the discretized unfolded KB contour only for the purpose of comparison to the reference data, which are obtained from the solution of the matrix equation \eqref{eq_dyson_matrix}. 

\begin{figure}[t]
\centering
\begin{minipage}{1\linewidth}
  \includegraphics[width=1\linewidth]{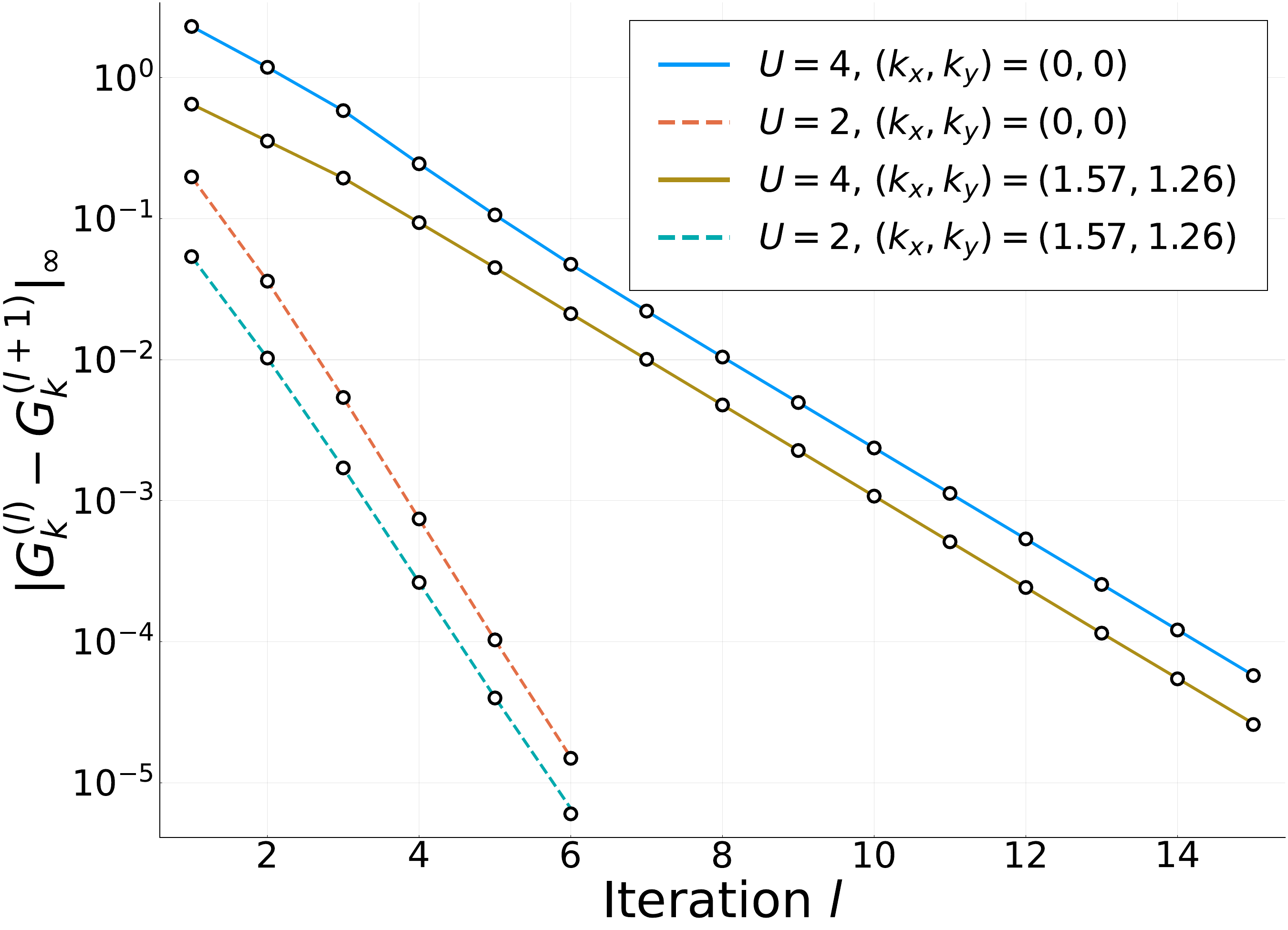}
  \end{minipage}
  \begin{minipage}{1\linewidth}
  \advance\leftskip0.1cm
  \includegraphics[width=0.99\linewidth]{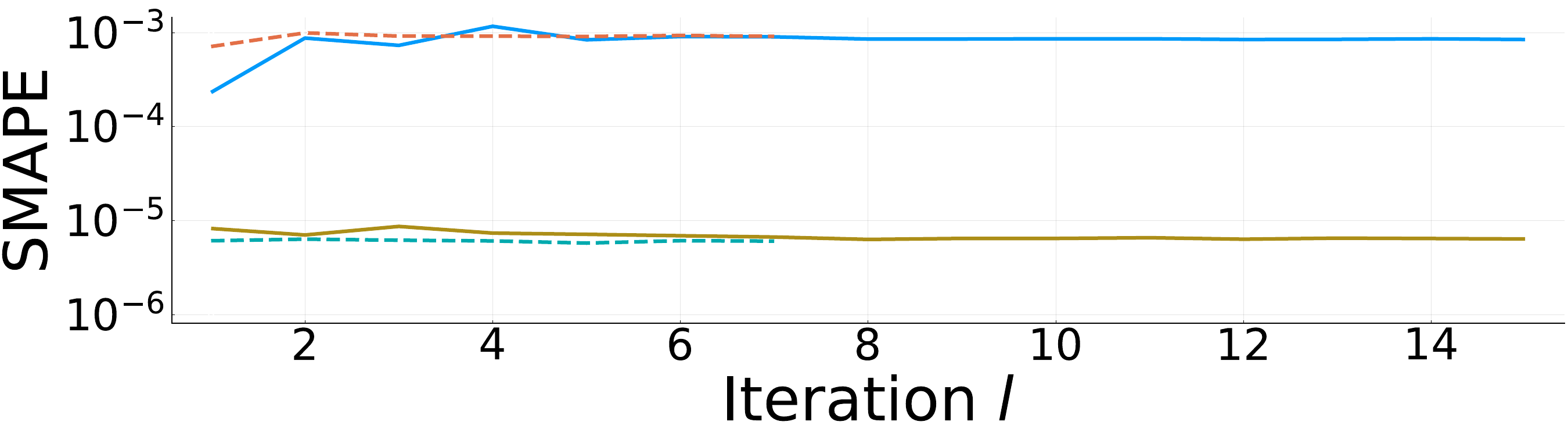}
  \end{minipage}
  \caption{Upper panel: Maximum norm error for $G_k^{(l)}-G_k^{(l+1)}$ as a function of iterations $l$ for $U=2$ and $4$, $\beta=2$, $(k_x, k_y) = (0,0)$ ($\Gamma$ point) and $(1.57, 1.26)$ (near the Fermi surface). 
  The lines are the result of the QTT implementation and the circles indicate the reference data from the matrix implementation.
  Lower panel: SMAPE for $G_\text{QTT}$  and $G_\text{matrix}$ as a function of iterations $l$, for the same parameters.
  }
  \label{fig_cauchy_error_eq}
\end{figure}

Figure~\ref{fig_cauchy_error_eq} illustrates the convergence of an equilibrium calculation in compressed form, and compares the results to the reference values from the non-compressed matrix calculation. These results are for the parameters $U=2$ and $4$, $\beta= 2$, $t_\text{max}=2$, $N_t = 400$, $ N_{\tau}= 220$ ($R=10$ binary digits, as $2(400+1)+(220+1)+1=1024=2^{10}$ \cite{footnote_plusone}), $N_k^2=20^2$, 
$\epsilon_\text{cutoff}=10^{-15}$ and maximum allowed bond dimension $D_{\text{max}}=120$. The top panel shows the difference $G_k^{(l)}-G_k^{(l+1)}$, with $l$ the iteration step, evaluated on the unfolded contour with the maximum norm $|\ldots|_\infty$ (maximum of the absolute values of the elements of the matrix). The solution can be considered as converged if this difference drops below a certain value $\epsilon$. For example, four significant digits corresponds to $\epsilon=10^{-4}$, since the Green's functions are of the order of unity. With the maximum norm, this accuracy is achieved after 6 (16) iterations for $U=2$ (4) and the two $k$-points presented in the figure. The lines in the figure show the results from the tensor train calculations, and the open circles those from the reference matrix calculation. The perfect agreement between the tensor train implementation and the matrix calculation demonstrates that there is no significant loss of accuracy by switching to the compressed representation. 

\begin{figure}[t]
\begin{minipage}{0.49\linewidth}
    \includegraphics[width=1\linewidth]{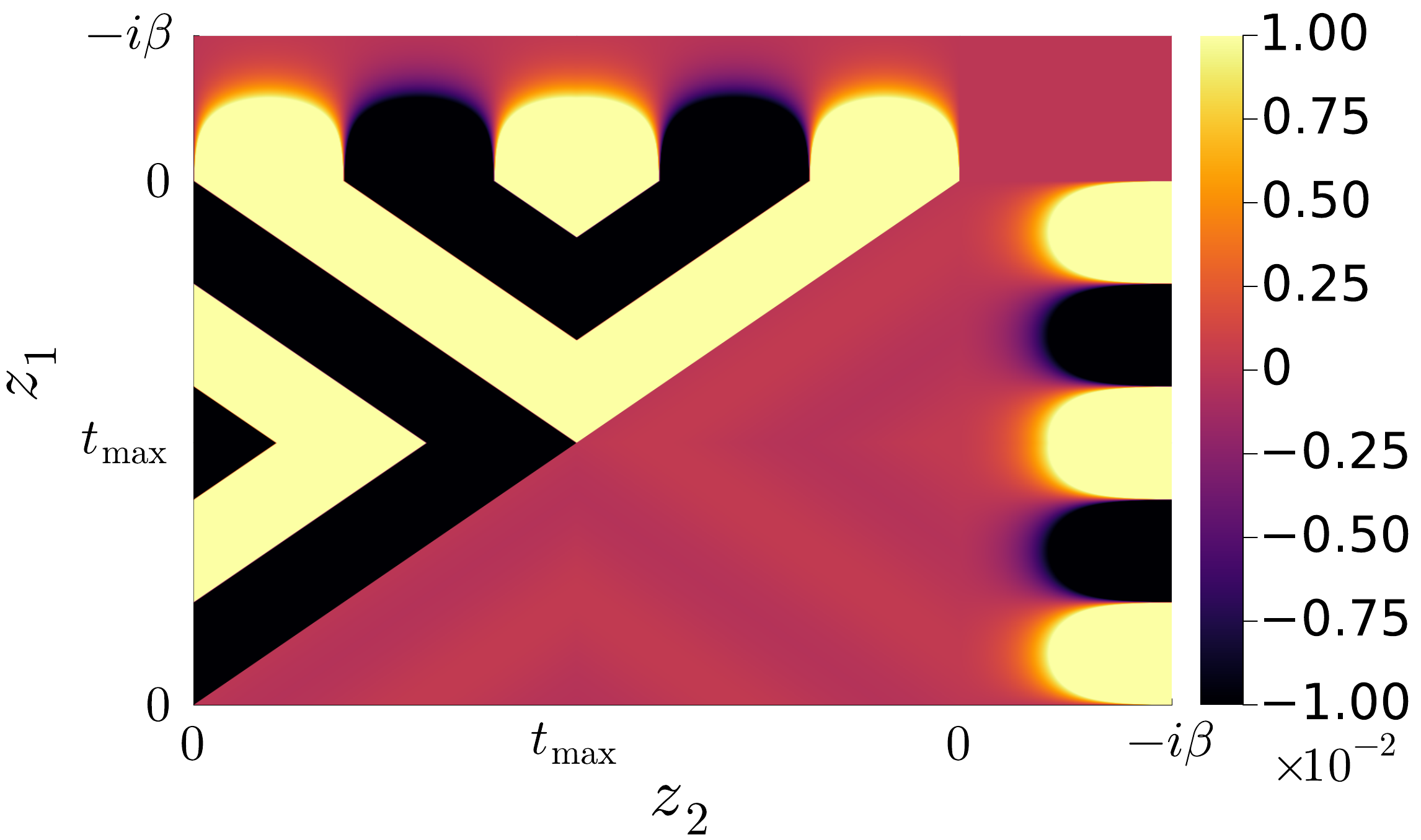}
\end{minipage}
\begin{minipage}{0.49\linewidth}
    \includegraphics[width=1\linewidth]{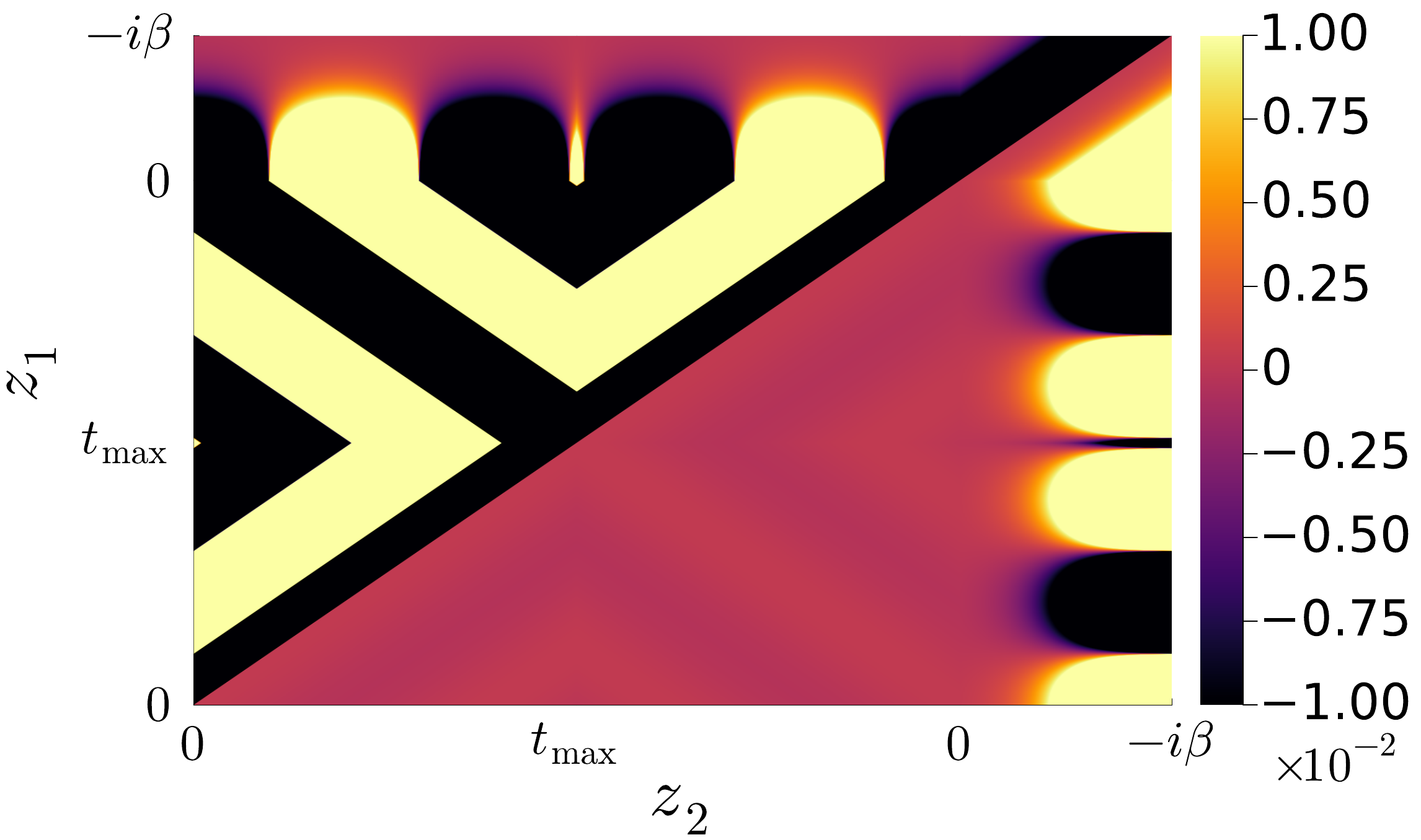}
\end{minipage}
\begin{minipage}{0.49\linewidth}
  \includegraphics[width=1\linewidth]{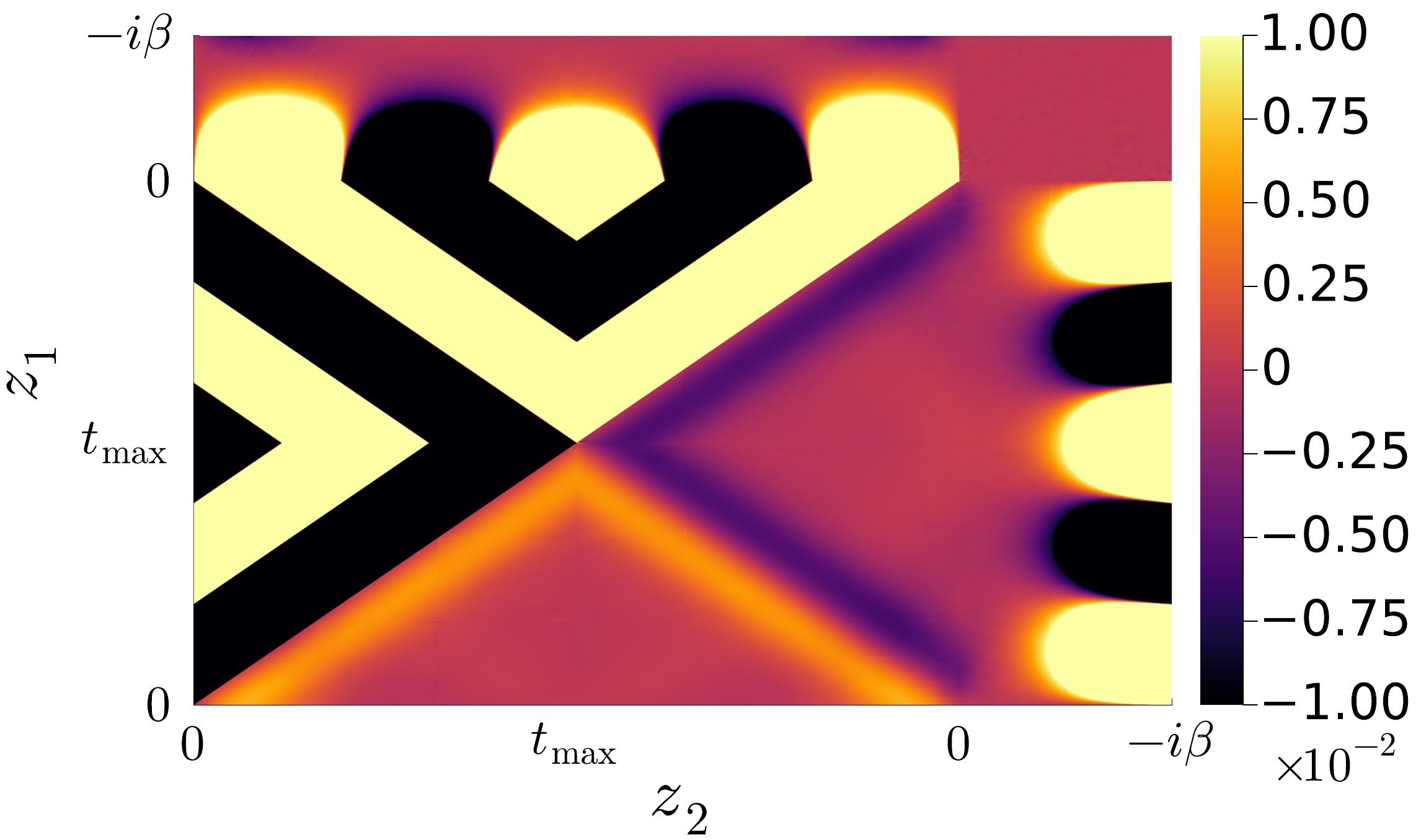}
\end{minipage}
\begin{minipage}{0.49\linewidth}
    \includegraphics[width=1\linewidth]{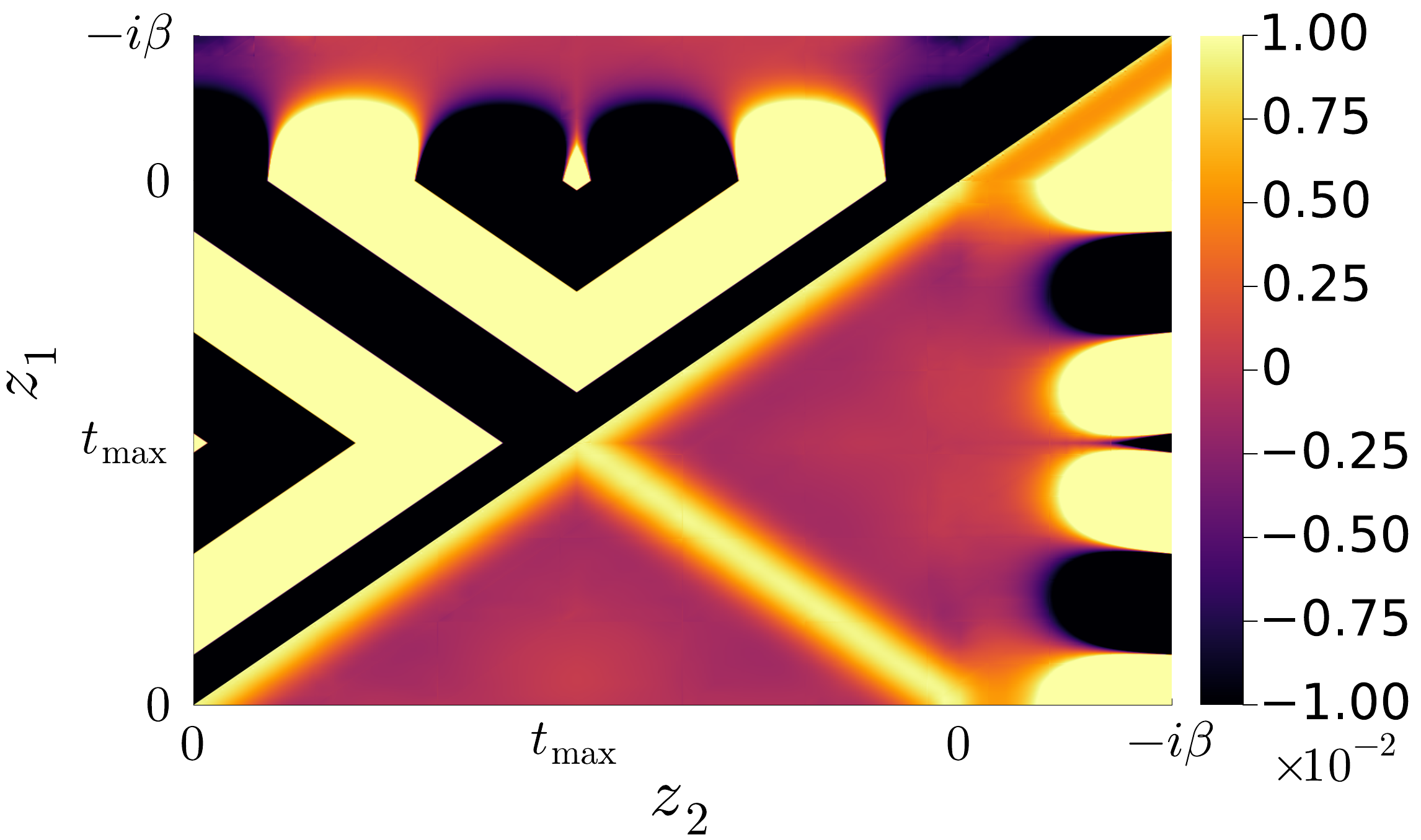}
\end{minipage}
\begin{minipage}{0.49\linewidth}
  \includegraphics[width=1\linewidth]{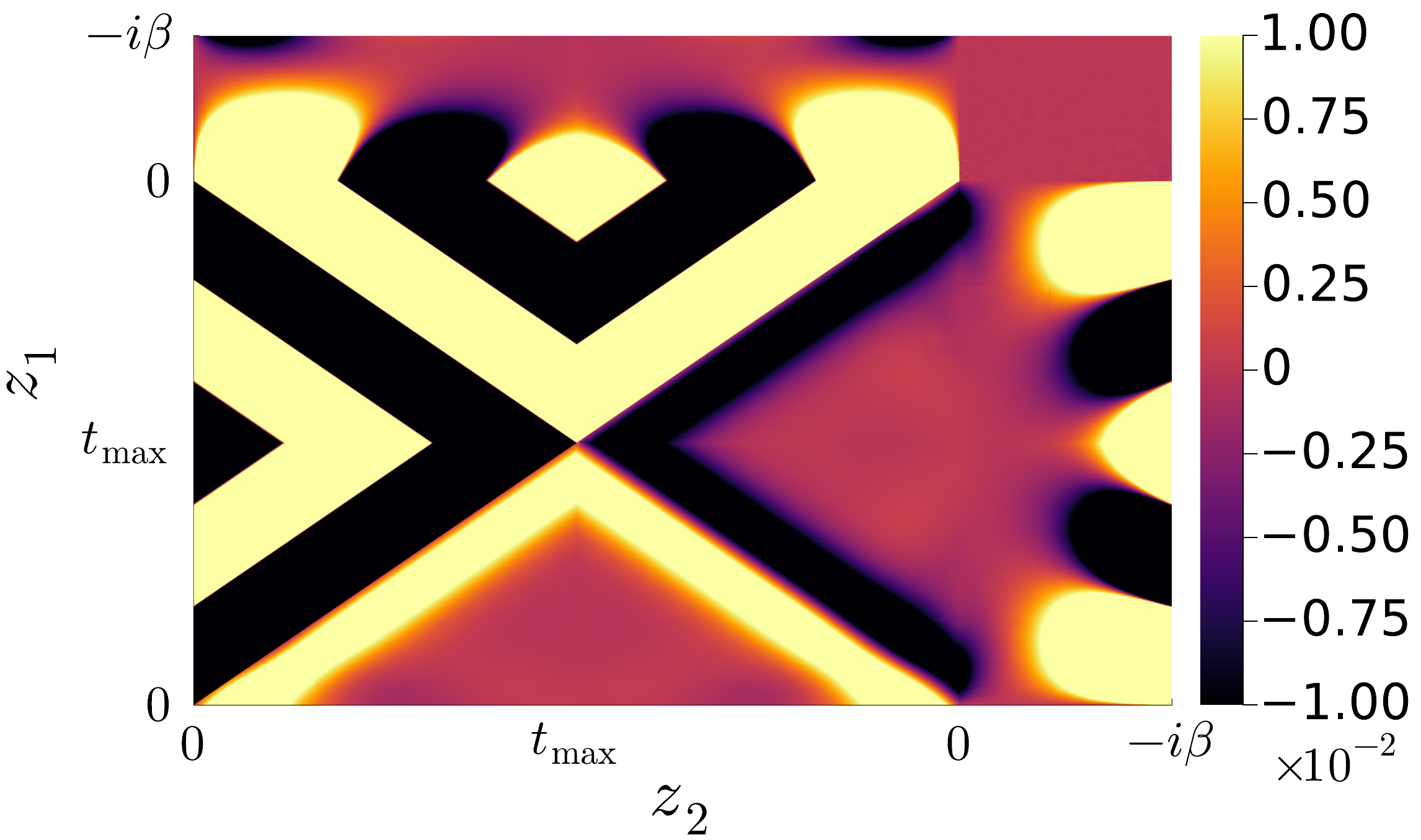}
\end{minipage}
\begin{minipage}{0.49\linewidth}
    \includegraphics[width=1\linewidth]{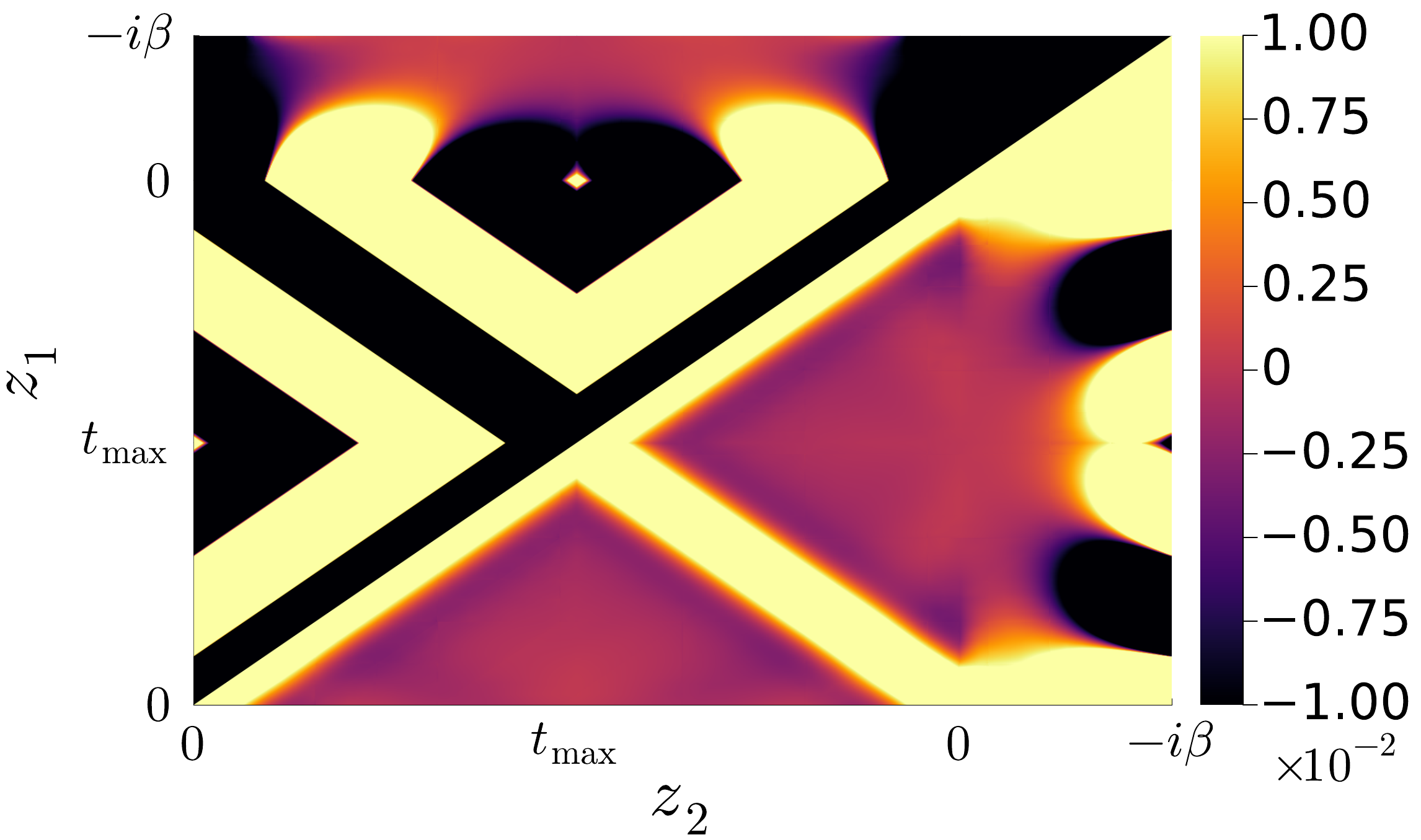}
\end{minipage}
  \caption{Real (left) and imaginary (right) noninteracting (top) and interacting (middle: $U=2$, bottom: $U=4$) Green's functions for $\beta=2$ and $k=(0,0)$. We restrict the color bar to the interval $[-0.01,0.01]$ to emphasize small structures.
  }
  \label{fig_G_eq}
\end{figure}

The maximum norm $|\ldots|_\infty$ is very sensitive to fluctuations in the difference between two Green's functions, and overemphasizes deviations which are confined to small regions in the two-time plane.  A global picture of the deviations between the Green's functions from the two methods is provided by the
SMAPE estimate defined in Eq.~(\ref{eq_smape}). This estimate yields a consistently small percentage error, independent of iteration number $l$, as shown in the lower panel of Fig.~\ref{fig_cauchy_error_eq}, which confirms that the two implementations produce essentially identical results. 

The speed of convergence does not depend strongly on the momentum $k$. On the other hand, for larger $U$, where the interacting Green's functions differ more from the noninteracting ones, the convergence slows down considerably. This could be potentially improved with dedicated mixing schemes, such as the Broyden method \cite{Zitko2009}. One should note, however, that self-consistent second order perturbation theory becomes unreliable for $U\gtrsim \text{bandwidth}/2=4$, so that the larger $U$ value shown in Fig.~\ref{fig_cauchy_error_eq} is at the upper end of the range of applicability. 

The real and imaginary parts of $G^0_k$ and the converged $G_k$ for $U=2$ and $4$ are plotted for $\beta=2$ and $k=(0,0)$ in Fig.~\ref{fig_G_eq}. As expected, the deviations from the noninteracting result (top panels) increase with increasing $U$. For a better visualization of small structures, we restrict the color bars to $[-0.01:0.01]$.

\subsection{CPU and RAM demand}\label{sec:CPU_RAM_demand}

\begin{figure}[t]
\begin{minipage}{1\linewidth}
  \includegraphics[width=1\linewidth]{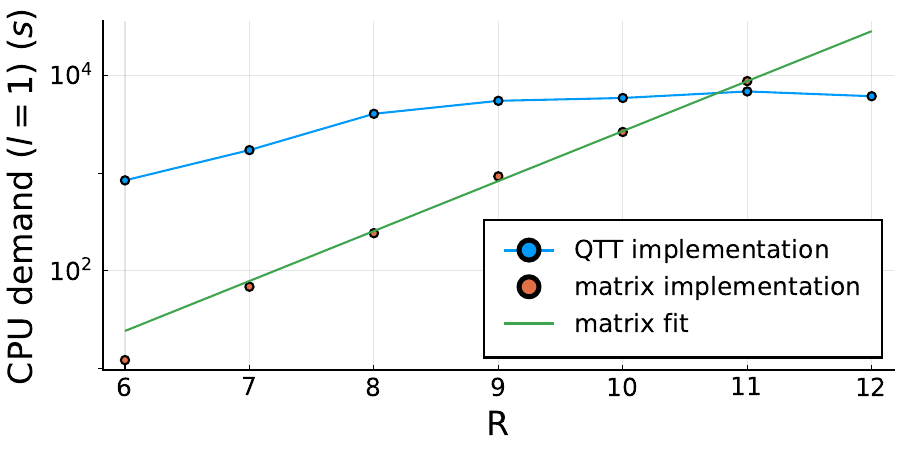}
\end{minipage}\\
\begin{minipage}{1\linewidth}
    \includegraphics[width=0.95\linewidth]{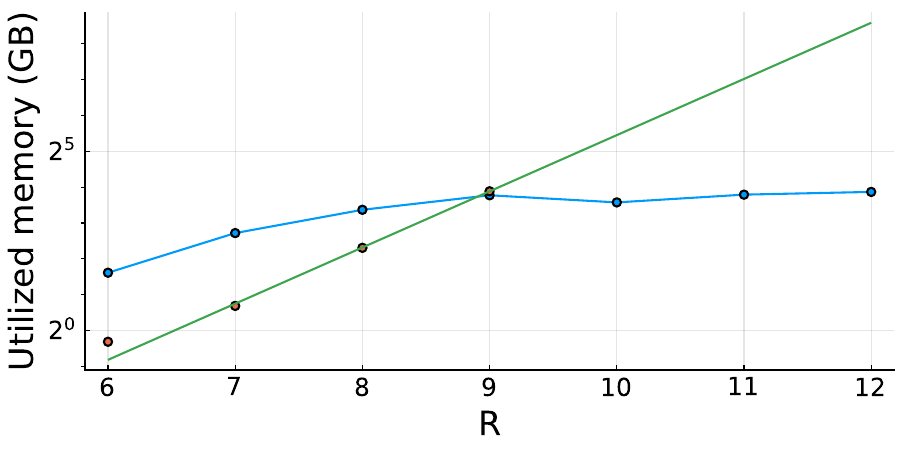}
\end{minipage}
  \caption{CPU and memory demand as a function of increasing mesh size for fixed $\beta=2$, $t_\text{max}=2$ (iteration $l=1$ for the CPU measurement). On the horizontal axis, we report the number of digits $R$ in the binary representation, which corresponds to $2^R$ time discretization points on the contour $\mathcal{C}$. Fitting the results for the matrix implementation to $(2^R)^b$ yields the exponent $b= 1.70\pm0.03$ $(1.57\pm 0.04)$ for the CPU (RAM) scaling.
  In the QTT calculation, we set the maximum bond dimension to $D=100$ and the cutoff to $\epsilon_{\text{cutoff}}=10^{-8}$.
  } 
  \label{fig_memory_CPU_demand}
\end{figure}
\begin{figure}
\begin{minipage}{1\linewidth}
  \includegraphics[width=1\linewidth]{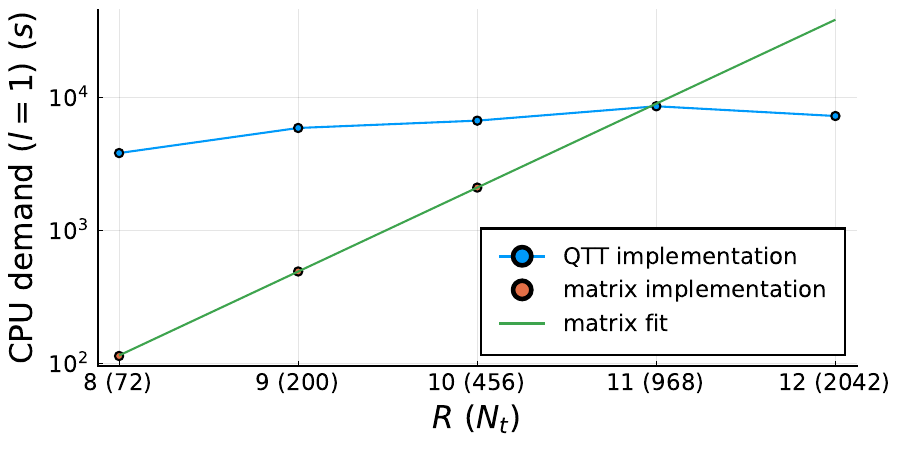}
\end{minipage}\\
\begin{minipage}{1\linewidth}
    \includegraphics[width=0.95\linewidth]{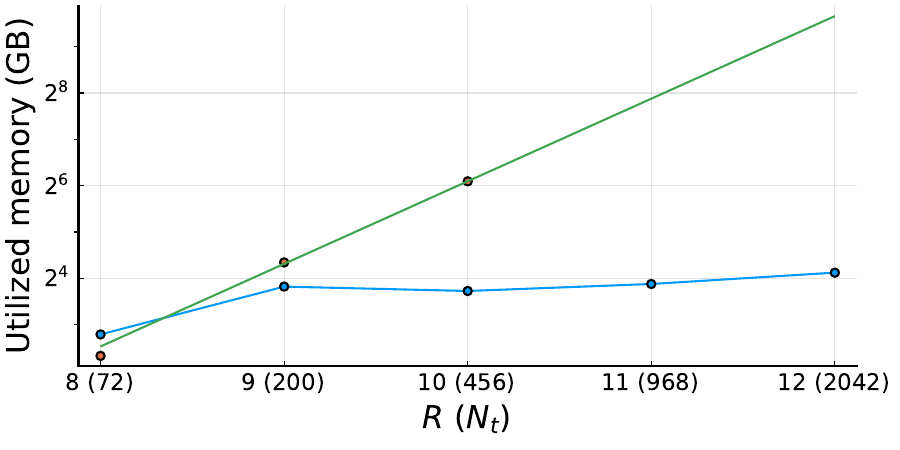}
\end{minipage}
  \caption{CPU and memory demand as a function of increasing number of digits $R$ in the binary representation.  The corresponding values of $N_t$ are indicated in the brackets on the horizontal axis ($t_\text{max}=N_t dt$). Here, $\beta=1$, $d\tau = 0.009$, and $N_\tau=108$ are fixed. Fitting to $(2^R)^b$ yields the exponent $b= 2.098\pm0.003$ $(1.78\pm 0.06)$ for the CPU (RAM) scaling.
  In the QTT calculation, we set the maximum bond dimension to $D=100$ and the cutoff to $\epsilon_{\text{cutoff}}=10^{-8}$.
  }
  \label{fig_memory_CPU_demand_t_max}
\end{figure}

The simulations were carried out on 128 Core AMD EPYC 7742 2.25 GHz processors with 
768 GB of random access memory (RAM)  using codes written in Julia 1.8.5. The QTT computations are implemented with the help of the \texttt{ITensors.jl} \cite{Fishman2022} library. We measure the CPU demand using the \textit{timed} function and report the time for the first iteration. The total physical RAM used is measured using the \textit{reportseff} Slurm command. 

In Fig.~\ref{fig_memory_CPU_demand} we show how the CPU and memory demand scales with the number of discretization steps for fixed $\beta=2$ and $t_\text{max}=2$, $U=2$ and $N_k^2=20^2$ ($N_k=20$ momentum points along each axis). In the case of the matrix calculations, the effort grows like a power-law of the matrix size, or exponentially $\sim (2^R)^b$ with increasing number of digits (per time variable) $R$ in the binary representation. 
Naively, one would expect that the memory demand grows quadratically ($b=2$) and the CPU time with the third power ($b=3$). The measured exponent for the memory demand is lower, because the matrices are still too small to fully dominate the RAM allocation. In the case of the CPU scaling, $b<3$ because our implementation of the Fourier transformation is rather inefficient, so that operations other than matrix multiplications account for a significant share of the CPU time. 

The QTT calculation, on the other hand, shows a saturation in both the CPU and memory demand beyond a certain value of $R$, which depends on the bond dimension. (Here, we set the maximum allowed bond dimension to $D=100$ and the cutoff to $\epsilon_{\text{cutoff}}=10^{-8}$.) Once all the physically relevant structures are fully resolved in the discretized form, the complexity of the QTT based calculation no longer increases by adding further digits (using a finer mesh), in contrast to the matrix calculation. As a result, even though the QTT implementation is not competitive for small time grids, it eventually outperforms the matrix implementation. 

One may be more interested in increasing $t_\text{max}$ with a fixed (small enough) time step $dt$, rather than increasing the number of discretization steps with fixed $t_\text{max}$. We performed a similar analysis with $dt= 0.005$, $d\tau = 0.009$, $\beta=1$, $N_\tau=108$ fixed in the QTT calculation.  We increase $R$ and adjust $N_t$ such that $2^R= (N_\tau+1) + (2N_t+2)+1$ \cite{footnote_plusone}. In the QTT calculations, we again limit the maximum bond dimension to a reasonable value, $D=100$, and set the cutoff to $\epsilon_{\text{cutoff}}=10^{-8}$. As shown Fig.~\ref{fig_memory_CPU_demand_t_max}, the CPU and memory demand shows a similar trend as reported in Fig.~\ref{fig_memory_CPU_demand}. In particular, the memory demand in the QTT calculation saturates, in contrast to the matrix implementation, where it increases almost quadratically with the total number of discretization steps ($\sim t_\text{max}$ for large $R$). 
The crossing point is between $R=8$ and $R=9$, which corresponds to a short time contour with $N_t<200$. 
The CPU demand in the QTT implementation also saturates and becomes lower than that of the matrix implementation for $R\ge 11$ ($N_t\gtrsim 1000$). 
Depending on the complexity of the function, it may become necessary though to increase $D$ with increasing $t_\text{max}$. 

\begin{figure}[t]
\centering
    \begin{minipage}{1\linewidth}
  \includegraphics[width=1\linewidth]{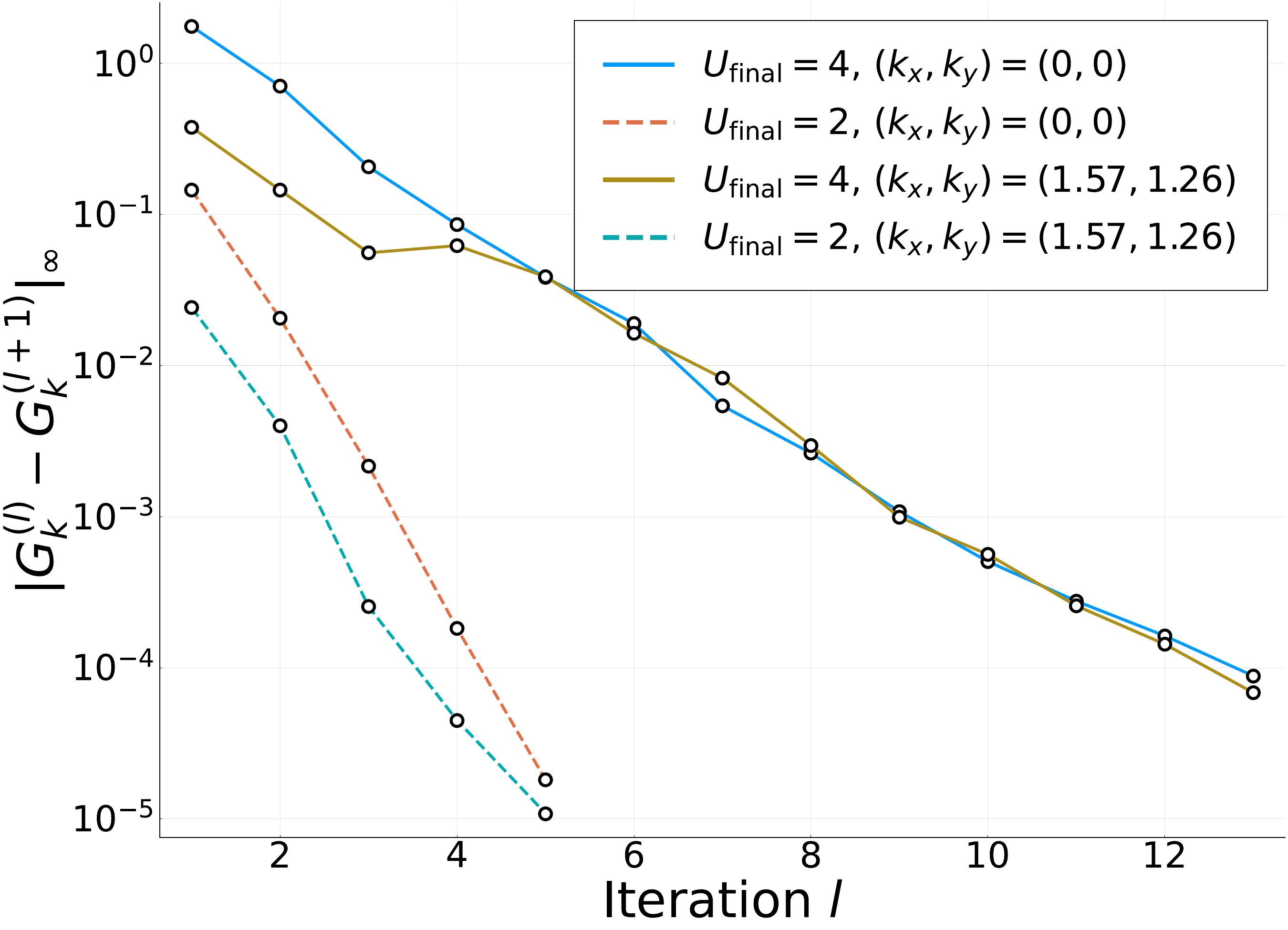}
    \end{minipage}
    \begin{minipage}{1\linewidth}
      \advance\leftskip0.1cm
  \includegraphics[width=0.99\linewidth]{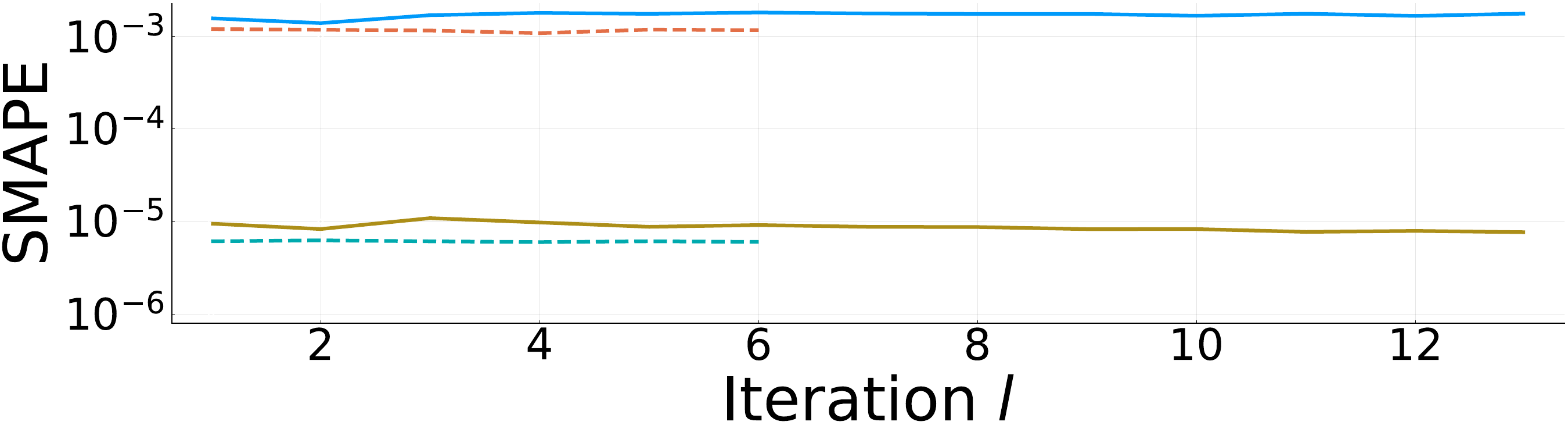}
  \end{minipage}
  \caption{
  Top panel: Maximum norm error for $G_k^{(l)}-G_k^{(l+1)}$ as a function of iterations $l$ for interaction ramps $U(t)$ to the indicated values of $U_\text{final}$, $\beta=2$ in the initial state, $(k_x, k_y) = (0,0)$ ($\Gamma$ point) and $(1.57, 1.26)$ (near the Fermi surface). The lines are the result of the QTT implementation and the circles indicate the reference data from the matrix implementation. Bottom panel: SMAPE for $G_\text{QTT}$  and $G_\text{matrix}$ as a function of iterations $l$, for the same parameters.  
  }
  \label{fig_cauchy_error_noneq}
\end{figure}

\subsection{Interaction ramp}

In this section, we show results for an interaction ramp calculation, starting from the noninteracting state. On the real-time axis, the interaction is ramped up as 
\begin{equation}
U(t) = \frac{U_\text{final}}{1+\exp(-\kappa (t-t_\text{ramp})/t_\text{max})}, 
\end{equation}
where $t_\text{ramp} = t_\text{max}/10$ and the steepness of the ramp is controlled by $\kappa=0.5$. The convergence of $G_k$ is illustrated for $U_\text{final}=2$ and $4$, initial $\beta=2$ and for the momenta $k=(0,0)$ and $(1.57,1.26)$ in Fig.~\ref{fig_cauchy_error_noneq}. 
Here, we use the same parameters as in Fig.~\ref{fig_cauchy_error_eq} ($\epsilon_{\text{cutoff}}=10^{-15}$ and $D_{\text{max}}=120$). 
The convergence behavior is similar to the equilibrium calculation (Fig.~\ref{fig_cauchy_error_eq}), but less monotonous in the case of $U=4$ and $k=(1.57,1.26)$. Again, the agreement between the QTT and matrix implementation is excellent, 
which confirms that also in nonequilibrium situations, the compression does not lead to any significant loss of accuracy 
As discussed previously, the maximum norm might detect some local fluctuations, which however do not represent a significant deviation between the QTT and matrix implementations. Indeed, SMAPE for $G_\text{QTT}$ and $G_\text{matrix}$ yields consistently low percentage errors for all iterations $l$, as shown in the lower panel of Fig.~\ref{fig_cauchy_error_noneq}. 
\begin{figure}[t]
\begin{minipage}{\linewidth}
  \includegraphics[width=1.\linewidth]{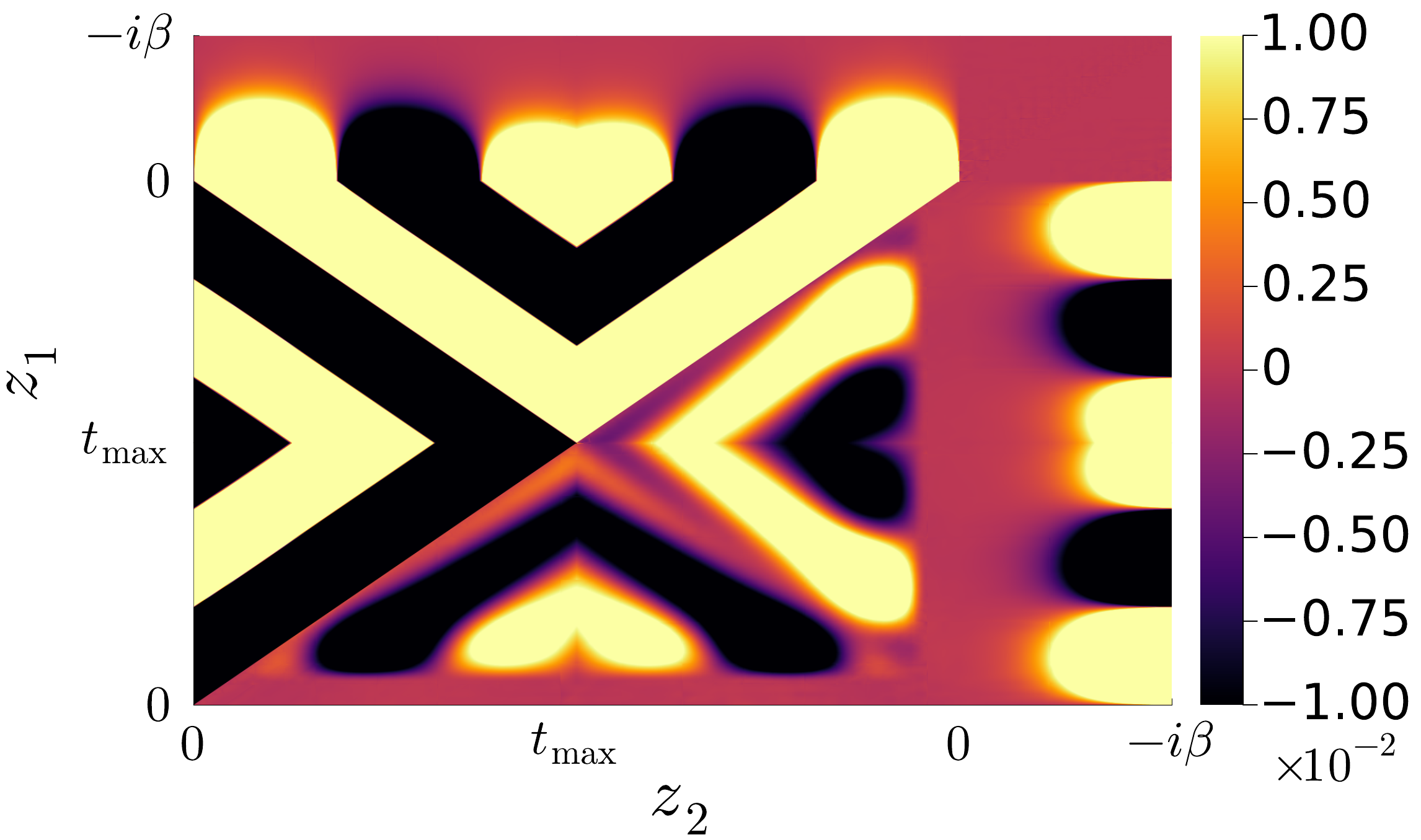}
\end{minipage}\\
\begin{minipage}{\linewidth}
    \includegraphics[width=1.\linewidth]{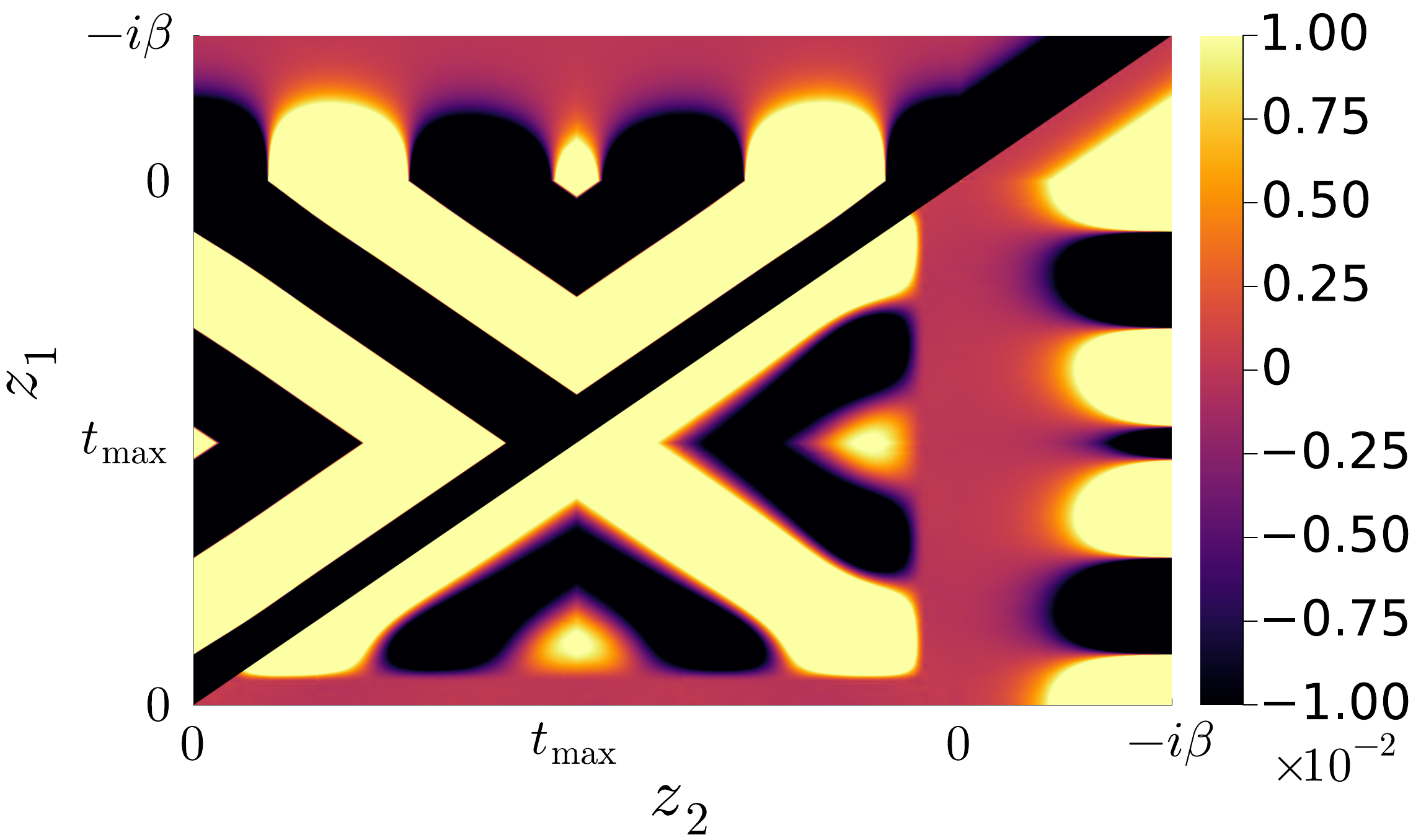}
\end{minipage}
  \caption{Real (top) and imaginary (bottom) part of the converged Green's function for the interaction ramp $U(t)$ to $U_\text{final}=4$, $\beta=2$ in the initial state, and $k=(0,0)$.
  }
  \label{fig_G_noneq}
\end{figure}

The real and imaginary parts of the converged $k=(0,0)$ Green's function, are shown in Fig.~\ref{fig_G_noneq} for the ramp to $U_\text{final}=4$. In contrast to the equilibrium results, this function now exhibits clearly non-time-translation-invariant features. For example, in the lesser component ($z_1\le z_2\le t_\text{max}$), the black area is no longer parallel to the diagonal $z_1=z_2$. 

\begin{figure}[t]
  \includegraphics[width=1\linewidth]{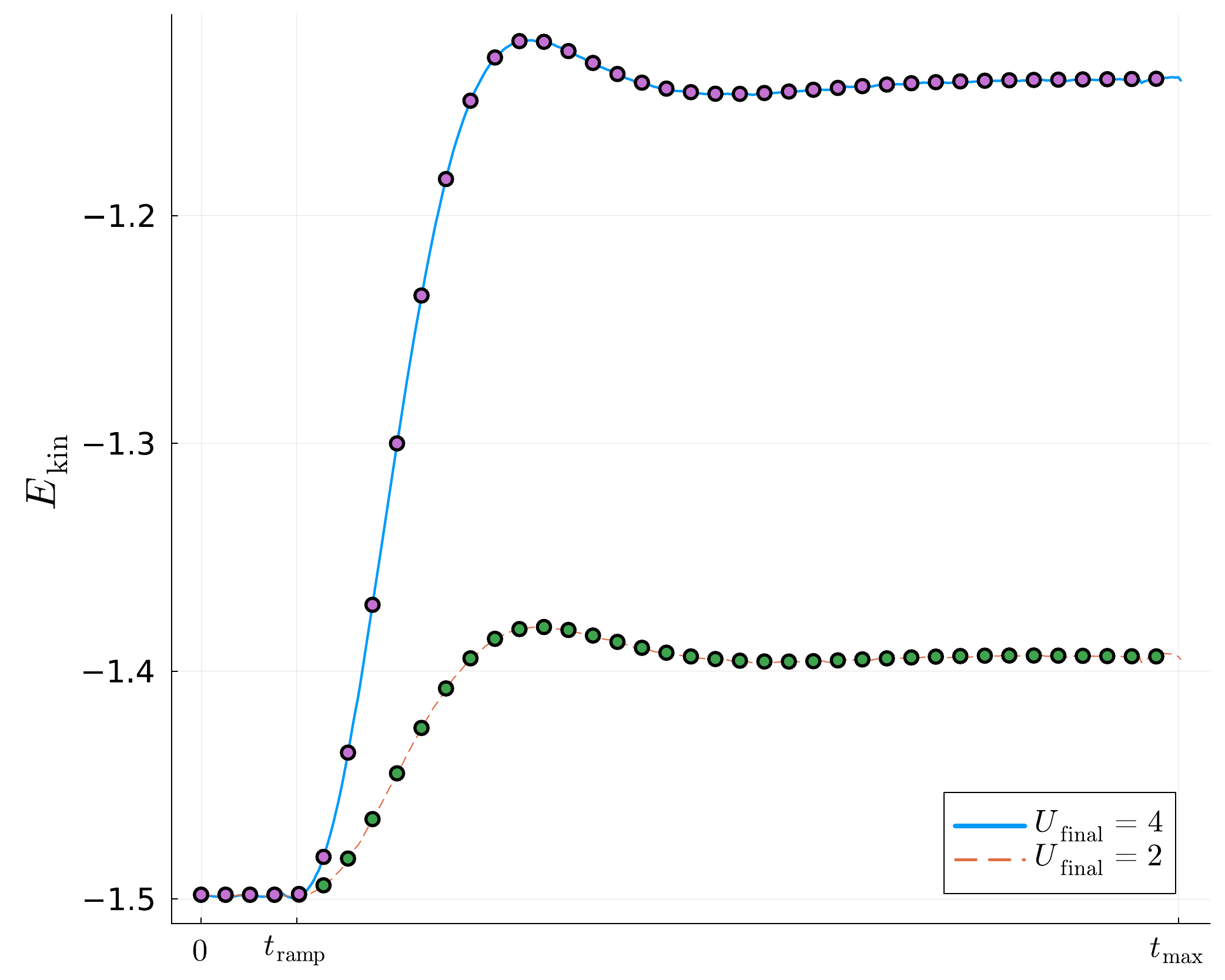}
  \caption{Kinetic energy of the lattice system subject to an interaction ramp $U(t)$ to $U_\text{final}=2$ and $4$. The lines (circles) show the results from the QTT (matrix) implementation. The initial inverse temperature is $\beta=2$.
  }
  \label{fig_kinetic_energy}
\end{figure}

In Fig.~\ref{fig_kinetic_energy}, we show the evolution of the kinetic energy per site
\begin{equation}
    E_\text{kin}(t) = \frac{-2i}{N_k^2}\sum_{{k}}\epsilon_{{k}}G_{{k}}^<(t,t).
\end{equation}
This energy contribution is negative in the initial equilibrium state, and increases during and after the ramp, due to the correlation induced band renormalization, and also due to heating. Once the correlated electronic structure of the interacting system is roughly established, the kinetic energy becomes approximately constant and approaches the thermalized value after strongly damped (overdamped) oscillations, as expected for a moderately correlated metallic system \cite{Eckstein2010b}. Also in the case of $E_\text{kin}(t)$, the results calculated in the QTT and matrix implementations agree, which demonstrates that realistic nonequilibrium simulations, including the calculation of relevant observables, can be implemented with compressed functions.

\section{Conclusions}
\label{sec:conclusions}

We demonstrated and tested the implementation of nonequilibrium Green's function based diagrammatic many-body calculations with QTT compressed two-time functions. Using self-consistent second order perturbation theory for the 2D Hubbard model as a simple but relevant application, we explained the implementation of the different calculation steps (Fourier transformation, scalar multiplication, element-wise product, sum and convolution) and used these routines to construct the second-order self-energy and to solve the lattice Dyson equation. In the present proof-of-principles study, we employed two-time functions defined on the unfolded KB contour, and restricted the QTT compression to the time dependence of these functions. To test and benchmark our calculations, we compared the QTT implementation to the matrix implementation with two-time functions defined on the discretized KB contour. 

Our investigation confirmed that the calculations with compressed objects reproduce the results from the matrix implementation up to high precision. An analysis of the CPU and RAM scaling revealed that the QTT implementation is not competitive with the matrix version for short time contours, but that it exhibits a more favorable scaling with increasing length of the time contour. For fixed $t_\text{max}$, the memory and CPU demands in the QTT implementation saturate once the number of digits in the binary representation is high enough that all relevant structures can be resolved. The QTT calculation is also not sensitive to $t_\text{max}$, as long as the maximum bond dimension needed for the accurate representation of the functions remains approximately constant. In practice, for the present model and implementation, the QTT calculation outperforms the matrix calculation for $R\ge 9$ (RAM) and $R\ge 11$ (CPU) or $t_\text{max}/dt \gtrsim 200$ and $t_\text{max}/dt \gtrsim 1000$, which are numbers of time points that are easily surpassed in realistic applications based on discretized contours.    

Since the computational effort for the relevant QTT operations scales steeply with the maximum bond dimension $D$, practical applications to (nonequilibrium) Green's function schemes should not employ the functions defined on the unfolded KB contour, but rather the lesser, retarded, left-mixing and Matsubara components \cite{Aoki2014,Nessi}, since this will allow to reduce $D$ by approximately a factor of 4. The latter approach also avoids ambiguities about the definition of the functions at $t=t'$. 

The QTT based approach is more naturally combined with a self-consistency loop which updates the function on the full time contour, than with a time-stepping scheme. For large $t_\text{max}$, the convergence properties of this approach will have to be further investigated.
Also, the dependence of the maximum bond dimension $D$ on the length of the contour needs to be studied in different relevant contexts, including quenches, periodically driven models, and systems with distinct characteristic timescales linked, e.~g., to prethermalization \cite{Berges2004} or nonthermal fixed points \cite{Tsuji2013}.  

It is possible that some form of coarse-graining, divide-and-conquer or patching will help to speed up the convergence.
Furthermore, this will reduce the bond dimension for each patch, and will allow efficient patch-wise massive parallelization.
A possible advantage of the divide-and-conquer QTT approach is that a given patch can be large, as long as its bond dimension stays reasonably small (e.g. $D=100$), while the time resolution is exponentially high with respect to $R$, with negligible discretization errors.
Another interesting direction for method development is the combination with tensor cross interpolation (TCI)~\cite{Fernandez2022, Ritter2023}.
The combination of quantics and TCI (QTCI)~\cite{Ritter2023} may accelerate the convolutions in the calculations of self-energies and the solution of Dyson equations.
Also, QTCI can be naturally combined with the divide-and-conquer approach.

A feature that distinguishes the QTT approach from the hierarchical low-rank matrix representation of Ref.~\cite{Kaye2021} is the possibility, at least in principle, to compress the dependence on momentum or orbital degrees of freedom by adding corresponding digits to the binary representation. If this can be done effectively, it would solve one of the major bottlenecks of nonequilibrium lattice simulations, namely the large memory cost for storing momentum-dependent two-time functions. 
We note that the aforementioned divide-and-conquer QTT approach can be regarded as a generalization of the hierarchical low-rank matrix representation:
The former uses a QTT with exponentially high resolution for each patch, while the latter uses a low-rank matrix decomposition with a fixed resolution. 

Systematic explorations of different patching approaches and multi-variable compression schemes are needed to gain more insights into the strengths and limitations of the various methods. 

\hspace{-1cm} 
\begin{acknowledgments}
The calculations were carried out on the Beo06 cluster at the University of Fribourg. We thank Y. Murakami for helpful discussions, and O. Simard for providing NESSi-based reference data. H.S. was supported by JSPS KAKENHI Grants No. 21H01041, No. 21H01003, and No. 23H03817 and JST PRESTO Grant No. JPMJPR2012, Japan.
\end{acknowledgments}


\begin{thebibliography}{99}
\bibitem{Giannetti2016} C. Giannetti, M. Capone, D. Fausti, M. Fabrizio, and F. Parmigiani, Ultrafast optical spectroscopy of strongly correlated materials and high-temperature superconductors: a non-equilibrium approach, Advances in Physics 65, 58 (2016). 
\bibitem{Sensarma2010} R. Sensarma, D. Pekker, E. Altman, E. Demler, N. Strohmaier, D. Greif, R. J\"ordens, L. Tarruell, H. Moritz, and T. Esslinger, Lifetime of double occupancies in the Fermi-Hubbard model, Phys. Rev. B {\bf 82}, 224302 (2010).
\bibitem{Berges2004} J. Berges, Sz. Borsanyi, and C. Wetterich, Prehermalization, Phys. Rev. Lett. {\bf 93}, 142002 (2004).
\bibitem{Tsuji2013} N. Tsuji, M. Eckstein, and P. Werner, Nonthermal Antiferromagnetic Order and Nonequilibrium Criticality in the Hubbard Model, Phys. Rev. Lett. {\bf 110}, 136404 (2013).
\bibitem{Stefanucci2013} G. Stefanucci and R. v. Leeuwen, Nonequilibrium Many- Body Theory of Quantum Systems: A Modern Introduction (Cambridge University Press, Cambridge, England, 2013).
\bibitem{Aoki2014} H. Aoki, N. Tsuji, M. Eckstein, M. Kollar, T. Oka, and P. Werner, Nonequilibrium dynamical mean-field theory and its applications, Rev. Mod. Phys. {\bf 86}, 779 (2014).
\bibitem{Bonitz2010} M. Bonitz and K. Balzer, Progress in Nonequilibrium Green’s Functions IV, Journal of Physics Conference Series {\bf 220}, 011001 (2010).
\bibitem{Eckstein2010} M. Eckstein and P. Werner, Nonequilibrium dynamical mean-field calculations based on the noncrossing approximation and its generalizations, Phys. Rev. B {\bf 82}, 115115 (2010).
\bibitem{Nessi} Michael Sch\"uler, D. Golez, Y. Murakami, N. Bittner, A. Hermann, Hugo U. R. Strand, P. Werner, and M. Eckstein, NESSi: The Non-Equilibrium Systems Simulation package, Computer Physics Communications {\bf 257}, 107484 (2020).
\bibitem{Lipavsky1986} P. Lipavsky, V. Spicka, and B. Velicky, Generalized Kadanoff-Baym ansatz for deriving quantum transport equations, Phys. Rev. B {\bf 34}, 6933 (1986).
\bibitem{Schueler2020} M. Sch\"uler, U. De Giovannini, H. H\"ubener, A. Rubio, M. A. Sentef, T. P. Devereaux, and P. Werner, How Circular Dichroism in time- and angle-resolved photoemission can be used to spectroscopically detect transient topological states in graphene, Phys. Rev. X {\bf 10}, 041013 (2020).
\bibitem{Schluenzen2020} N. Schl\"unzen, J.-P. Joost, and M. Bonitz, Achieving the Scaling Limit for Nonequilibrium Green Functions Simulations, Phys. Rev. Lett. {\bf 124}, 076601 (2020).
\bibitem{Pavlyukh2022} Y. Pavlyukh, E. Perfetto, D. Karlsson, R. van Leeuwen, and G. Stefanucci, Time-linear scaling nonequilibrium Green's function methods for real-time simulations of interacting electrons and bosons. I. Formalism, Phys. Rev. B {\bf 105}, 125134 (2022).
\bibitem{Schueler2018} M. Sch\"uler, M. Eckstein, and P. Werner, Truncating the memory time in nonequilibrium DMFT calculations, Phys. Rev. B {\bf 97}, 245129 (2018).
\bibitem{Stahl2022} C. Stahl, N. Dasari, J. Li, A. Picano, P. Werner, and M. Eckstein, Memory truncated Kadanoff-Baym equations, Phys. Rev. B {\bf 105}, 115146 (2022).
\bibitem{Kaye2021} J. Kaye and D. Golez, Low Rank Compression in the Numerical Solution of the Nonequilibrium Dyson Equation, SciPost Phys. {\bf 10}, 091 (2021).
\bibitem{Shinaoka2023} H. Shinaoka, M. Wallerberger, Y. Murakami, K. Nogaki, R. Sakurai, P. Werner, and A. Kauch, Multiscale Space-Time Ansatz for Correlation Functions of Quantum Systems Based on Quantics Tensor Trains, Phys. Rev. X {\bf 13}, 021015 (2023).
\bibitem{Schollwock2011} U. Schollw\"ock, The density-matrix renormalization group in the age of matrix product states, Annals of physics {\bf 326}, 96 (2011).
\bibitem{Cirac2021} J. I. Cirac, D. Perez-Garcia, N. Schuch, and F. Verstraete, Matrix product states and projected entangled pair states: Concepts, symmetries, theorems, Rev. Mod. Phys. {\bf 93}, 045003 (2021).
\bibitem{Stoudenmire2010} E. M. Stoudenmire, S. R. White, Minimally entangled typical thermal state algorithms,
New J. Phys. 12 055026 (2010)
\bibitem{Fishman2022} Matthew Fishman, Steven R. White, E. Miles Stoudenmire, The ITensor Software Library for Tensor Network Calculations,
SciPost Phys. Codebases 4 (2022).
\bibitem{footnote_plusone} We add a column and a row of zeros to match the power of 2.
\bibitem{Zitko2009} R. Zitko, Convergence acceleration and stabilization of dynamical mean-field theory calculations, Phys. Rev. B {\bf 80}, 125125 (2009).
\bibitem{Langreth1976} D. C. Langreth, Linear and Nonlinear Electron Transport in Solids, edited by J. T. Devreese and V. E. van Doren (Plenum
Press, New York, 1976).
\bibitem{Eckstein2010b} M. Eckstein, M. Kollar, and P. Werner, Interaction quench in the Hubbard model: Relaxation of the spectral function and the optical conductivity, Phys. Rev. B {\bf 81}, 115131 (2010).
\bibitem{Fernandez2022} Y. N. Fern\'andez, M. Jeannin, P. T. Dumitrescu, T. Kloss, J. Kaye, O. Parcollet, and X. Waintal, Learning Feynman Diagrams with Tensor Trains, Phys. Rev. X {\bf 12}, 041018 (2022).
\bibitem{Ritter2023} M. K. Ritter, Y. N. Fernández, M. Wallerberger, J. von Delft, H. Shinaoka, X. Waintal, Quantics Tensor Cross Interpolation for High-Resolution, Parsimonious Representations of Multivariate Functions in Physics and Beyond, arXiv:2303.11819v1 (to appear in PRL).

\end{thebibliography}
\end{document}